\documentclass[10pt,journal,compsoc]{IEEEtran}

\usepackage{graphicx}
\usepackage{subfigure}

\usepackage{amsmath,bm}
\usepackage{amssymb}
\usepackage{amsthm}
\usepackage{epsfig,algorithmic,algorithm}

\usepackage[english]{babel}

\usepackage{times}
\usepackage{arydshln}
\usepackage{multirow}

\ifCLASSOPTIONcompsoc
  \usepackage[nocompress]{cite}
\else
  \usepackage{cite}
\fi
\ifCLASSINFOpdf

\else

\fi

\begin{document}
\title{Addressing the Item Cold-start Problem by Attribute-driven Active Learning}

\author{Yu~Zhu, Jinhao Lin, Shibi He,
        Beidou Wang, Ziyu Guan, Haifeng Liu
        and~Deng~Cai,~\IEEEmembership{Member,~IEEE}
\IEEEcompsocitemizethanks{\IEEEcompsocthanksitem Y. Zhu, J. Lin, S. He and D. Cai are with State Key Laboratory of CAD\&CG, College of Computer Science, Zhejiang University, Hangzhou, China, 310027. E-mail: \{zhuyu\_cad, fenixl\}@zju.edu.cn, frankheshibi@gmail.com, dcai@zju.edu.cn.
\IEEEcompsocthanksitem B. Wang is with School of Computing Science, Simon Fraser University, Canada. E-mail: beidouw@sfu.ca.
\IEEEcompsocthanksitem Z. Guan is with the College of Information and Technology, Northwest University of China, Xi'an, China 710127. E-mail: ziyuguan@nwu.edu.cn.
\IEEEcompsocthanksitem H. Liu is with the College of Computer Science, Zhejiang University, Hangzhou, China, 310027. E-mail: haifengliu@zju.edu.cn.}
\thanks{Manuscript}} 

\markboth{Journal of \LaTeX\ Class Files}
{Shell \MakeLowercase{\textit{et al.}}: Bare Advanced Demo of IEEEtran.cls for Journals}

\IEEEtitleabstractindextext{%
\begin{abstract}
In recommender systems, cold-start issues are situations where no previous events, e.g. ratings, are known for certain users or items. In this paper, we focus on the item cold-start problem. Both content information (e.g. item attributes) and initial user ratings are valuable for seizing users' preferences on a new item. However, previous methods for the item cold-start problem either 1) incorporate content information into collaborative filtering to perform hybrid recommendation, or 2) actively select users to rate the new item without considering content information and then do collaborative filtering. In this paper, we propose a novel recommendation scheme for the item cold-start problem by leverage both active learning and items' attribute information. Specifically, we design useful user selection criteria based on items' attributes and users' rating history, and combine the criteria in an optimization framework for selecting users. By exploiting the feedback ratings, users' previous ratings and items' attributes, we then generate accurate rating predictions for the other unselected users. Experimental results on two real-world datasets show the superiority of our proposed method over traditional methods.
\end{abstract}

}

\maketitle

\IEEEdisplaynontitleabstractindextext

\IEEEpeerreviewmaketitle

\ifCLASSOPTIONcompsoc
\IEEEraisesectionheading{\section{Introduction}\label{sec:introduction}}
\else
\section{Introduction}
\label{sec:introduction}
\fi

\IEEEPARstart{R}{ecommender} systems (RS) have become extremely common in recent years, and are applied in a variety of domains, from virtual community web sites like movielens.org to electronic commerce companies like amazon.com. In spite of the widespread application of RS, one difficult and common problem is the cold-start problem, where no prior events, like ratings or clicks, are known for certain users or items. The user cold-start problem may lead to the loss of new users due to the low accuracy of recommendations in the early stage. The item cold-start problem may make the new item miss the opportunity to be recommended and remain ``cold'' all the time. In this paper, we focus on the item cold-start problem, where recommendations are required for items that no one has yet rated.
\begin{figure}[tb!]
\begin{center}
\includegraphics[scale=0.39]{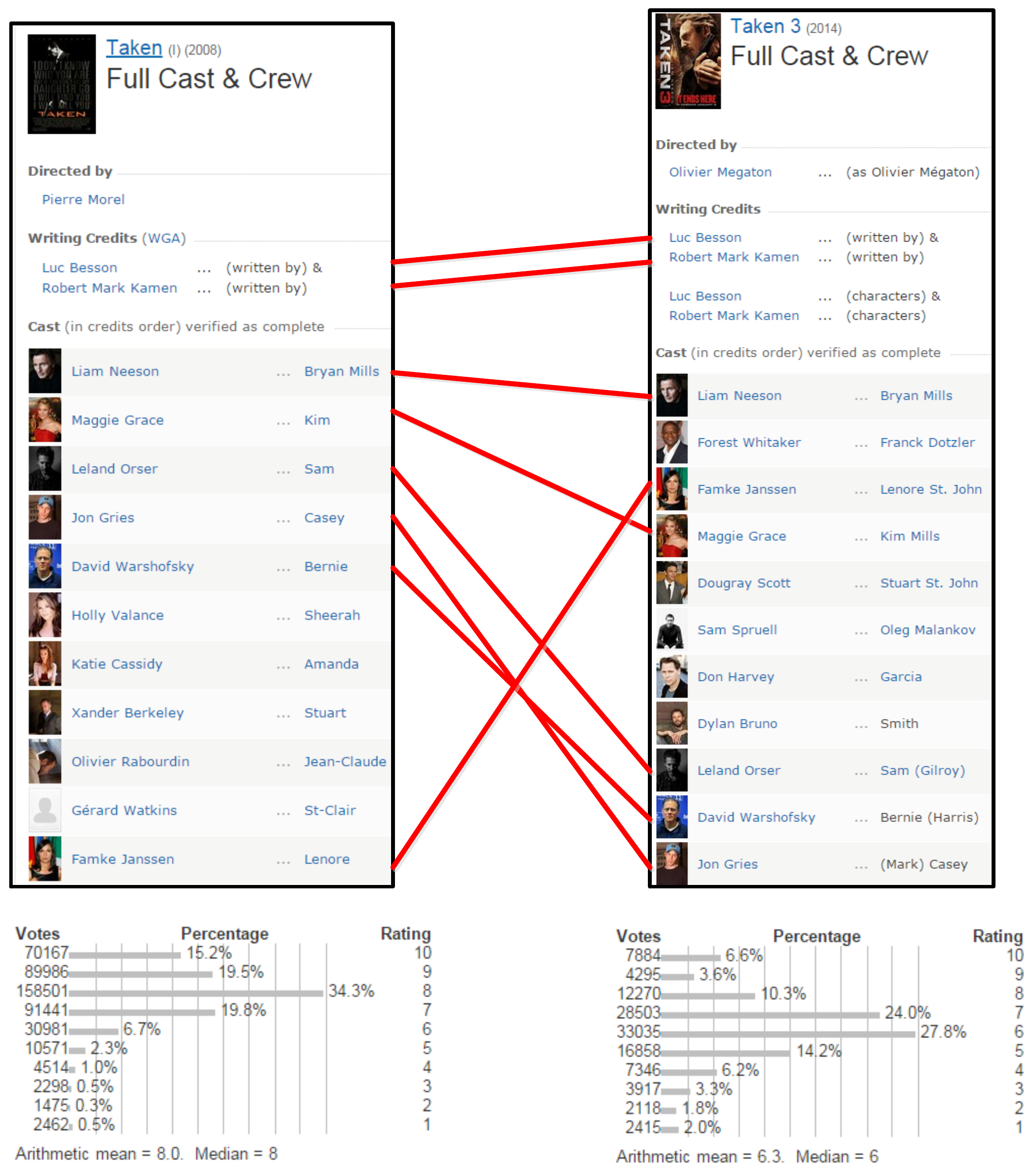}
\end{center}
\vspace*{-10pt}
   \caption{Attribute information and users' rating distributions of films \emph{Taken} and \emph{Taken 3}. Genres of these two films are both \emph{Action} and \emph{Thriller}. Common scriptwriters and actors are connected by red lines. The overall ratings of \emph{Taken} are high with a mean equal to 8.0 and the mean rating is 6.3 for \emph{Taken 3}. Attributes are modeled as features in this paper, whose values are 1 if corresponding attributes exist or 0 if they do not exist. In this example, features include all genres, directors, scriptwriters and actors.
}
\label{fig:takenDis}
\vspace*{-10pt}
\end{figure}

Content information, such as item attributes, were exploited to address such issues in previous methods \cite{hong:2013co,gantner:2010learning,hauger:2008comparison}. However, items with similar attributes may be of different interest for the same user. As shown in Figure \ref{fig:takenDis} (data is collected from \emph{IMDB} \footnote{http://www.imdb.com/}), movie \emph{Taken} is favored by many people after release, with a mean rating equal to 8.0. When the follow-up \emph{Taken 3} was first released in 2014, it can be seen as a ``cold'' film. Since genres, screenwriters and many actors of these two films are the same, then if we exploit film attributes to perform hybrid recommendations, we may recommend this ``cold'' film to users who favored \emph{Taken} before. However, as can be seen from the figure, the peak of \emph{Taken 3}'s overall ratings moves down to rating 6, which means that many users might favor \emph{Taken} but would give low ratings to \emph{Taken 3}. One reason could be that, although \emph{Taken 3} and \emph{Taken} have many attributes in common, \emph{Taken 3} has a lower quality than \emph{Taken}, thus users who favored \emph{Taken} before may dislike \emph{Taken 3}, i.e. the recommendation of \emph{Taken 3} to users who favored \emph{Taken} before might not be accurate. Therefore, it is not a safe way to handle the cold-start issue based on film attributes only. A natural solution is to select a small set of users to watch this ``cold'' film first, whose feedback can give us more understanding of users' preferences on this ``cold'' film. Then we can perform more accurate recommendations. Interestingly, this is similar to the key idea of active learning in the machine learning literature \cite{rubens:2015active}.

Most works that apply active learning to recommender systems focus on the user cold-start problem \cite{elahi:2014active,rashid:2008learning,golbandi:2011adaptive}. New users' preferences are typically obtained by directly interviewing the new user about what his interest is, or asking him to rate several items from carefully constructed seed sets. Seed sets may be constructed based on popularity, contention and coverage \cite{rashid:2008learning}. Items in these constructed seed sets will be rated by every new user. However, the item cold-start problem is different because items cannot be interviewed and typically there are no users willing to rate every new item. Thus we need to construct different user sets to rate different new items, ensuring that users are not always selected for rating requests. In addition, the user set for each new item must be carefully constructed so that we can learn as much as possible about the new item given a limited number of rating requests. However, limited works have been conducted to address the item cold-start problem by active learning. \cite{aharon:2015excuseme,anava:2015budget} use the active learning idea but ignore items' attribute information. Meanwhile, they select users based on limited criteria. In fact, the new item's attributes give us some understanding of this item and can be exploited to improve our user selection strategy. For example, we tend to select users who favor attributes existing in the new item, since these users are more willing to give ratings.

In this paper, we propose a novel recommendation framework for the item cold-start problem, where items' attributes are exploited to improve active learning methods in recommender systems. The attribute-driven active learning scheme has following characteristics:
\begin{itemize}
\item Explicitly distinguishing 1) whether a user will rate the new item and 2) what rating the user will give to the new item. The former helps us to select users who are willing to give ratings to the new item (feedback ratings). The latter allows us to exploit the rating distribution to improve the selection strategy. For example, we expect to select users who give diverse ratings to generate unbiased predictions. This is easy to understand since if we select users who all give high ratings, then the trained prediction model will generate high biased ratings for all other users, though the other users may not favor the new item at all.
\item Personalized selection strategy to ensure fairness. We construct our selection strategy based on four criteria, of which two are personalized criteria. The personalized criteria ensure that for new items with different attributes, users selected by our method would be different. This can avoid selecting the same user to rate every new item, which will negatively influence the user experience. These criteria are uniformly modeled as an integer quadratic programming (IQP) problem, which can be efficiently solved by some relaxation.
\item Dynamic active learning budget. In previous active learning works \cite{huang:2007selectively,anava:2015budget}, the \emph{budget} of active learning (i.e. the number of users selected for rating requests) for a new item is fixed. However, in real-world applications, 1) some new items are under the attention of a small set of users (not popular), e.g. films with unpopular actors and directors, and 2) some would be obviously favored by almost all users (popular and not controversial), e.g. \emph{Harry Potter and the Deathly Hallows: Part 2} \footnote{http://www.imdb.com/title/tt1201607/}, while 3) others are popular but controversial, and the recommender is not sure about users' preferences on them, e.g. although \emph{Taken 3} is famous, the qualities of films previously acted by its main actors vary a lot, thus it is difficult to predict users' preferences on \emph{Taken 3}. It is the items in the third case that need more feedback ratings so as to be learned more about. In this paper, we are the first to propose a dynamic active learning budget so that the limited active learning resources will be properly distributed, which can improve the overall prediction accuracy.
\item Considering \emph{exploitation}, \emph{exploration} and their trade-off. Traditional active learning methods aim at maximizing the performance measured on unselected instances in the prediction phase \cite{chattopadhyay:2012batch,chakraborty:2015batchrank}, regardless of the cost in the active learning phase, since they assume the labeling cost for each instance is the same. However, in our active learning phase, we prefer a rating request for a user who is willing to rate the item rather than a user who is not, because the latter one will negatively influence the user experience. Our solutions are inspired by \cite{rubens:2007influence, rokach:2008pessimistic}, which try to maximize the sum of \emph{rewards} by balancing the trade-off of \emph{exploitation} and \emph{exploration}. The \emph{rewards} in our task contain two parts, i.e. the user experience in the active learning phase and the prediction phase, respectively. By exploiting ``existing knowledge'' (\emph{exploitation}) from the model trained in Figure 3 (b), we are able to select willing users to obtain good user experience in the active learning phase. For the user experience in the prediction phase, we want to learn as much ``new knowledge'' (\emph{exploration}) about unselected users' preferences as possible, so as to generate accurate rating predictions for them. Note that users selected that best satisfy \emph{exploitation} may not be the most helpful for \emph{exploration}. Therefore, the ``exploitation-exploration trade-off'' in our task lies in how we optimize our user selection strategy, in order to obtain relatively good user experience in both of the active learning phase and the prediction phase. Our method considers both of these two goals and can further balance their trade-off by adjusting the parameter setting.

\end{itemize}

\section{Related Work}
\subsection{The Item Cold-start Problem}
To address the item cold-start problem, a common solution is to perform hybrid recommendations by combining content information and collaborative filtering \cite{agarwal:2009regression,gunawardana:2008tied,park:2009pairwise,nasery:2016recommendations}. A regression-based latent factor model is proposed in \cite{agarwal:2009regression} to address both cold and warm item recommendations in the presence of items' features. Items' latent factors are obtained by low-rank matrix decomposition. \cite{park:2009pairwise} solve a convex optimization problem, instead of the matrix decomposition, to improve this work. Another approach based on Boltzmann machines is proposed in \cite{gunawardana:2008tied, gunawardana:2009unified} to solve the item cold-start problem, which also combines content and collaborative information. LCE \cite{saveski:2014item} exploits the manifold structure of the data to improve the performance of hybrid recommendations. Other works are under a different setting where few ratings of new items exist, but no items' attribute information is known. \cite{aharon:2012dynamic, aizenberg:2012build} use a linear combination of raters' latent factors weighted by their ratings to estimate new items' latent factors.

\subsection{Active Learning in Recommender Systems}
Most active learning methods in recommender systems focus on the user cold-start problem, where they select items to be rated by newly-signed users \cite{rubens:2015active,elahi:2016survey}. We briefly introduce these methods since most of them can also be adapted to our new item task. The Popularity strategy \cite{golbandi:2010bootstrapping,golbandi:2011adaptive} and the Coverage strategy \cite{golbandi:2010bootstrapping} are two representative attention-based methods, where the former one selects items that have been frequently rated by users and the latter one selects items that have been highly co-rated with other items. Uncertainty reduction methods aim at reducing the uncertainty of rating estimates \cite{golbandi:2010bootstrapping,rubens:2007influence,rokach:2008pessimistic}, model parameters \cite{hofmann:2003collaborative, jin:2004bayesian} and decision boundaries \cite{danziger:2007choosing}. Error reduction methods try to reduce the prediction error on the testing set by either 1) optimizing the performance measure (e.g. minimizing \emph{RMSE}) on the training set \cite{golbandi:2010bootstrapping,golbandi:2011adaptive}, or 2) directly controlling the factors that influence the prediction error on the testing set \cite{rubens:2009output, settles:2008multiple}. \cite{harpale:2008personalized} uses some initial ratings to perform personalized active learning in a non-attribute context. There are also combined strategies \cite{rubens:2007influence,mello:2010active,elahi:2012adapting} considering several objectives at the same time. When applied to our new item task, some of these works require a few initial ratings on new items, which are not available in our task. The other works do not need initial ratings, but perform active learning regardless of the content information. However, the new item's content information gives us some understanding of the new item and we can exploit it to better perform active learning. In addition, methods such as the Popularity strategy \cite{golbandi:2010bootstrapping,golbandi:2011adaptive} and the Coverage strategy \cite{golbandi:2010bootstrapping} always select the same set of users, which negatively influence the user experience.

\cite{anava:2015budget,aharon:2015excuseme} are works which also address the item cold-start problem in an active learning scheme. However, they both focus on the pure collaborative filtering model and do not consider the content information either.

\subsection{The Exploitation-exploration Trade-off}
Some works also consider the exploitation-exploration trade-off \cite{rubens:2007influence, rokach:2008pessimistic}. Many of the promising solutions come from the study of the multi-armed bandit problem \cite{feldman:2015recommendations}. The key idea of these solutions is to simultaneously optimize one's decisions based on existing knowledge (i.e. \emph{exploitation}) and new knowledge which would be acquired through these decisions (i.e. \emph{exploration}), in order to maximize the sum of rewards earned through a sequence of actions. The $\epsilon$-Greedy algorithm \cite{watkins:1989learning} selects the arm which has the best estimated mean reward with probability $1 - \epsilon$, and otherwise randomly selects an other arm. UCB-like (UCB refers to Upper Confidence Bound) algorithms \cite{auer:2002using, dani:2008stochastic, abbasi:2011improved} firstly calculate the confidence bound of all arms and then select the arm with the largest upper confidence bound. The insight is that arms with a large mean reward (exploitation) and high uncertainty (exploration) would have large upper confidence bound. Thompson Sampling algorithms \cite{thompson:1933likelihood, agrawal:2012analysis, russo:2014learning} firstly calculate the probability distribution of the mean reward for each arm, then draw a value from each distribution and finally select the arm with the largest drawn value.

Our task and the multi-armed bandit problem share some common features, e.g. both considering the exploitation-exploration trade-off. However, they have some key differences. In the multi-armed bandit problem, the arms are selected one by one and a reward is generated immediately after each arm is selected. Hence, many solutions (e.g. UCB-like algorithms, Thompson Sampling algorithms) design their selecting strategies based on previous rewards. However, in our setting, a batch of users are selected at the same time, without knowing other users' feedback (reward), thus many solutions for the multi-armed bandit problem cannot be applied to our task.

\section{preliminaries and model}
\subsection{Task Definition and Solution Overview}
\begin{figure}[htb]
\begin{center}
\includegraphics[scale=0.5]{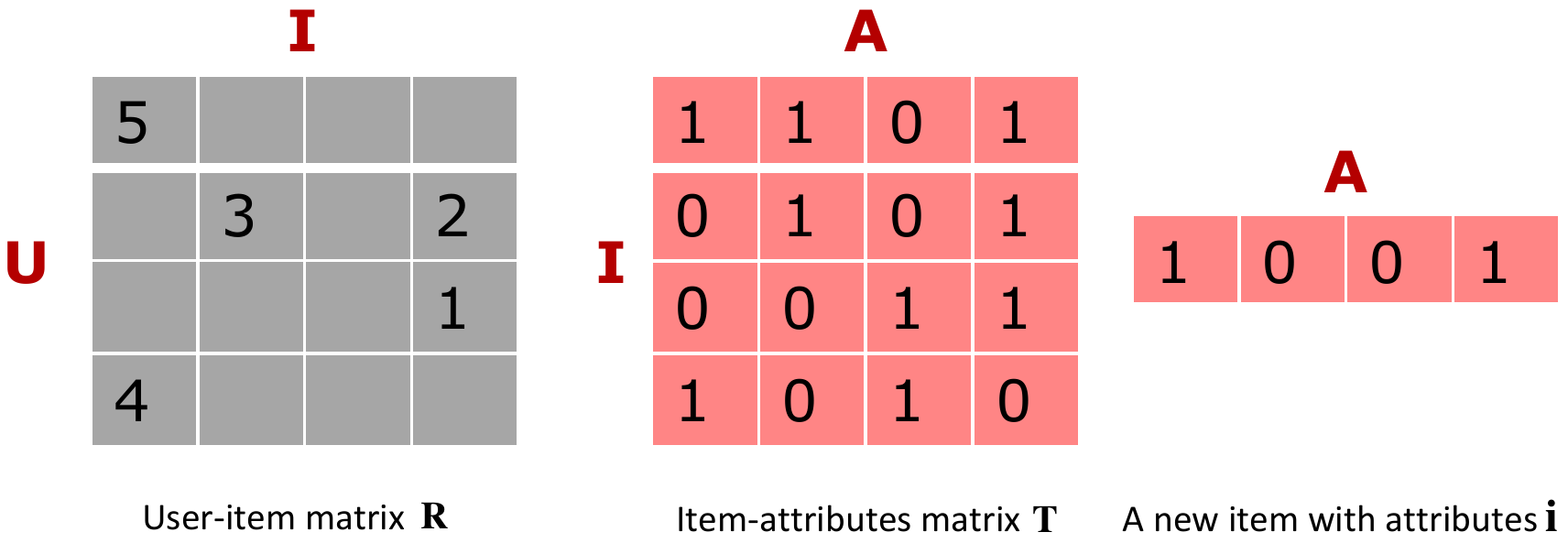}
\end{center}
\vspace*{-10pt}
   \caption{Representations of user-item matrix $\mathbf{R} \in R^{|U| \times |I|}$, item-attributes matrix $\mathbf{T} \in R^{|I| \times |A|}$ and the vector $\mathbf{i} \in R^{1 \times |A|}$ representing a new item with attributes.
}
\label{fig:matrix representation}
\vspace*{-10pt}
\end{figure}

We use $U$, $I$ and $A$ to denote the users, items and attributes set, respectively. Our task is: given a user-item rating matrix $\mathbf{R} \in R^{|U| \times |I|}$, an item-attribute matrix $\mathbf{T} \in R^{|I| \times |A|}$ and a new item $i_{new}$, whose attributes are denoted as a vector $\mathbf{i} \in R^{1 \times |A|}$, predict users' ratings on the new item $predict\_rating(u,i_{new}), u\in U$. $\mathbf{R}$, $\mathbf{T}$ and $\mathbf{i}$ are shown in Figure \ref{fig:matrix representation}. In this paper, the task is solved via two phases. The first one is the active learning phase, which is to solve: which users should be selected to rate $i_{new}$, so as to learn about $i_{new}$ as much as possible. The second one is the prediction phase, which is to solve: once given selected users' feedback, how to accurately predict the other users' ratings on $i_{new}$. For the active learning phase, we carefully select users based on four useful criteria, which involve both classification tasks and regression tasks. For the prediction phase, we model it as a pure regression task. We use Factorization Machines \cite{rendle:2012factorization} to model all classification tasks and regression tasks in these two phases.
\subsection{Factorization Machines}
Factorization Machines(FM) is a state-of-art framework for latent factor models which can incorporate rich features. Here we briefly introduce it. Please refer to \cite{rendle:2012factorization} for a more detailed description. The prediction problem is described by a matrix $\mathbf{X} \in R^{N\times D}$ and a vector $\mathbf{y}\in R^{N\times 1}$, where each row $\mathbf{x}\in R^{1\times D}$ of $\mathbf{X}$ is one instance with $D$ real-valued features and each entry $y$ in $\mathbf{y}$ is the label of one instance. FM can model nested feature interactions up to an arbitrary order between the $D$ features of $\mathbf{x}$. For feature interactions up to $2$-order, they are modeled as:
\begin{equation}\label{eq:FM}
\begin{aligned}
y^\prime(\mathbf{x}) = &w(0) + \sum_{i=1}^D w(i)x(i)\\
&+\sum_{i=1}^D\sum_{j=i+1}^D x(i)x(j)\sum_{f=1}^k V(i,f)V(j,f),
\end{aligned}
\end{equation}
where $k$ is the dimensionality of the factorization and the model parameters $\mathbf{\Theta} = \{w(0),\mathbf{w},\mathbf{V}\}$ are: $w(0)\in R, \mathbf{w}\in R^{D\times 1}, \mathbf{V}\in R^{D\times k}$. $w(i)$, $x(i)$ and $V(i,f)$ are entries of $\mathbf{w}$, $\mathbf{x}$ and $\mathbf{V}$, respectively. The form of FM (Equation (\ref{eq:FM})) is very general and can be applied to many applications.

In our task, we need to predict 1) what rating a user will give to an item and 2) whether a user will rate an item.

For the regression task to predict what rating a user will give to an item, features can contain users, items and attributes of items, and labels are ratings. The first term of the right-hand side in Equation (\ref{eq:FM}) is a bias of the system. If $w(0)$ is large, then there is a bias towards high ratings, which may be due to the good user experience of the system design. The second term is a bias of unary features, that is, some optimistic users tend to give high ratings to every item, and some popular items or items with popular attributes always gain high ratings. The last term is a bias of feature interactions. Many users only give high ratings to certain items (or items with certain attributes) which they are really interested in.

For the classification task to predict whether a user will rate an item, the analysis of each term in Equation (\ref{eq:FM}) is similar, except that labels now represent whether users will rate items or not.

\section{Our method}
\subsection{Select Users to Rate the New Item}
As described in section 3.1, we first need to carefully select users to rate the new item $i_{new}$, so as to learn about $i_{new}$ as much as possible. Users are selected based on the following four criteria.

\begin{figure*}[htb!]
\begin{center}
\subfigure[Representations of $\mathbf{R}$, $\mathbf{R}^{01}$ and $\mathbf{T}$]{
\includegraphics[scale=0.45]{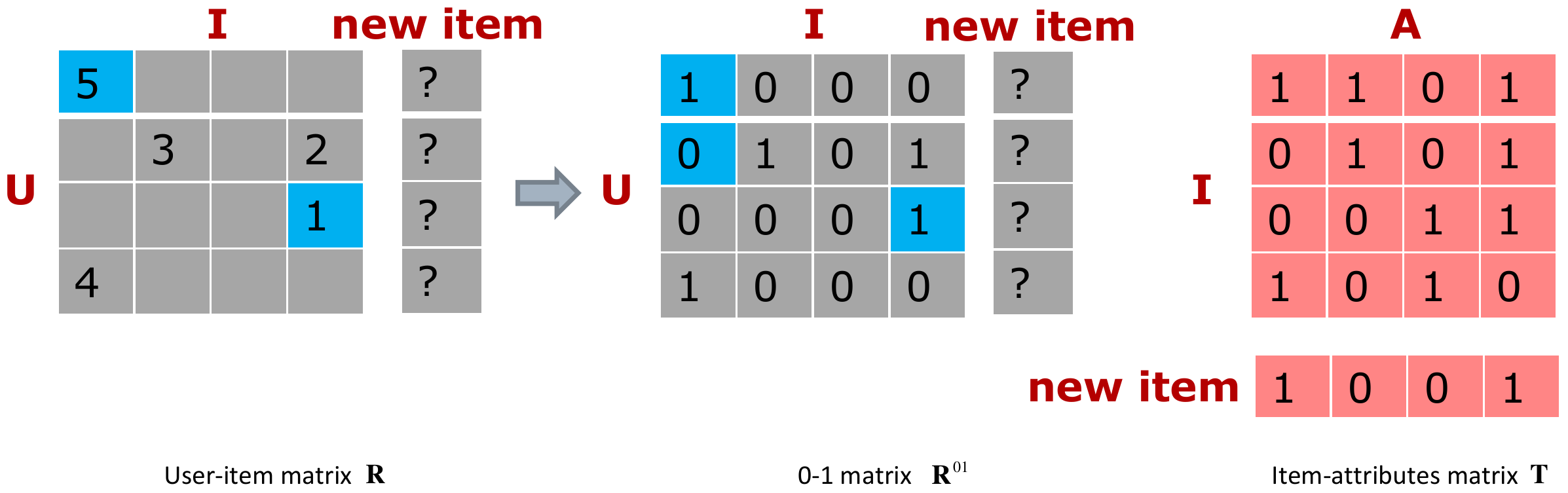}
}
\subfigure[Factorization Machines to model whether users will rate items]{
\begin{minipage}[b]{0.43\textwidth}
\centering
\includegraphics[scale=0.3]{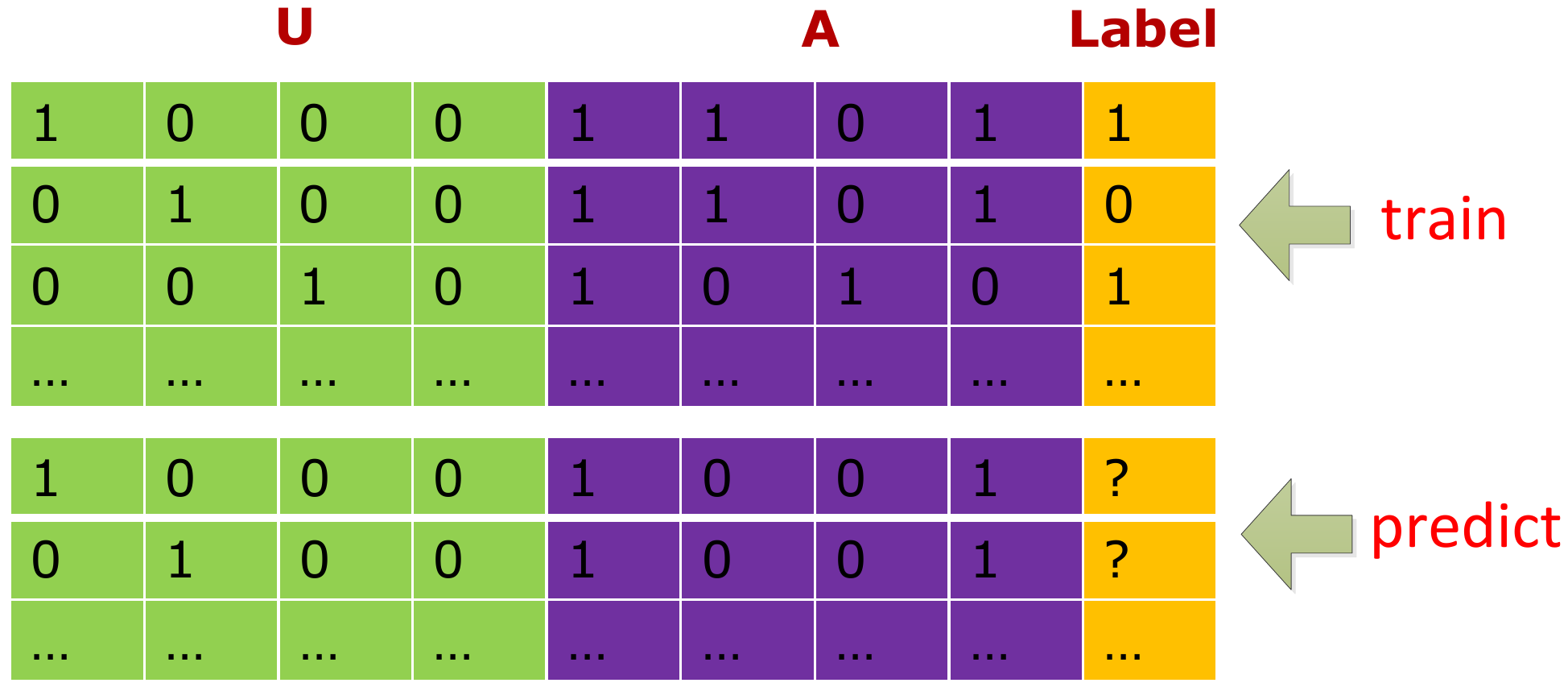}
\end{minipage}
}
\subfigure[Factorization Machines to model what ratings users will give to items ]{
\begin{minipage}[b]{0.43\textwidth}
\centering
\includegraphics[scale=0.3]{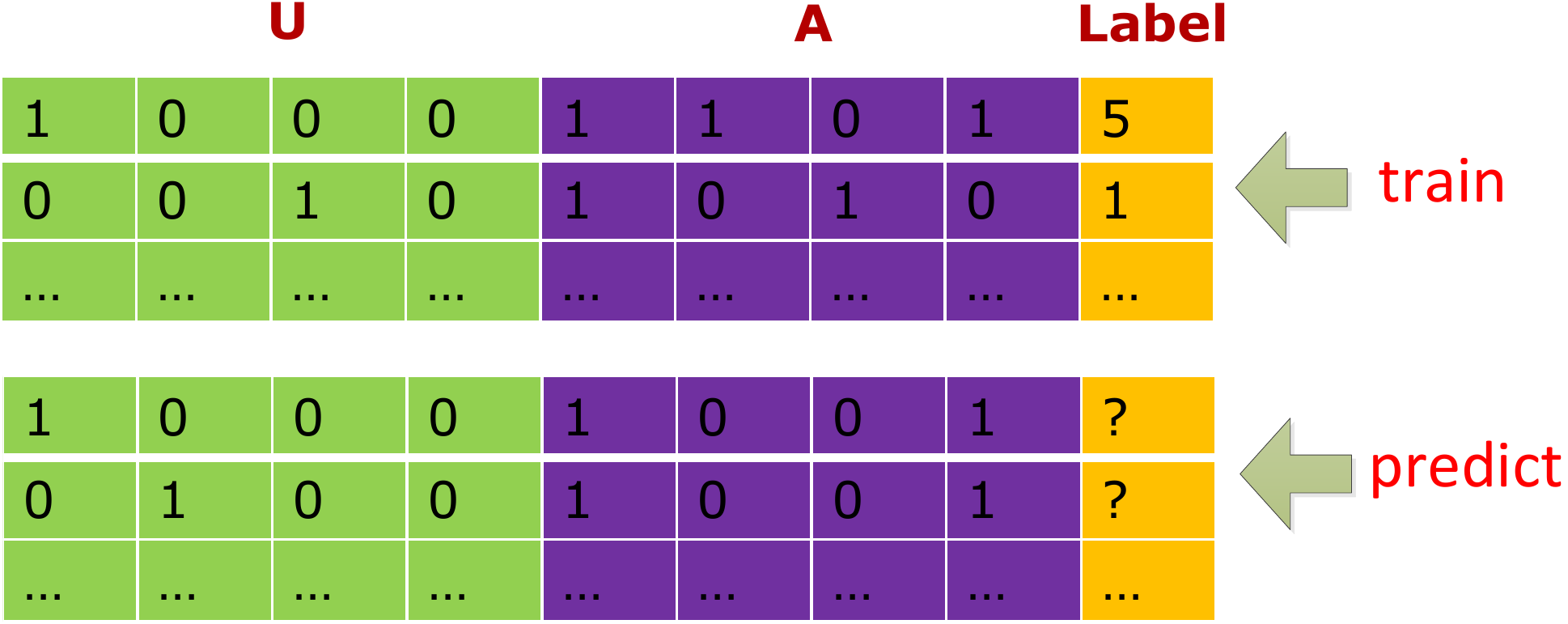}
\end{minipage}
}
\end{center}
\vspace*{-10pt}
   \caption{(a) shows how $\mathbf{R}$ is transformed to $\mathbf{R}^{01}$ \cite{koren:2008factorization}. (b) is the classification model. Each row is an instance, which corresponds to an entry of $\mathbf{R}^{01}$. For example, the first row corresponds to an entry of $\mathbf{R}^{01}$ with row 1 and column 1. Row 1 indicates the first user, thus the feature set labeled as `U' is (1,0,0,0) (one-hot representation). Column 1 indicates the first item, thus the feature set labeled as `A' is equal to the first row of $\mathbf{T}$, i.e. attributes of the first item. The first three rows correspond to $\mathbf{R}^{01}$'s three blue entries in (a). For testing, each user is predicted to show whether he will rate the new item. (c) is the regression model. Each row is an instance, which corresponds to an entry of $\mathbf{R}$. The first two rows correspond to $\mathbf{R}$'s two blue entries in (a). For testing, each user is predicted to show what rating he will give to the new item.
}
\label{fig:chooseToRate}
\vspace*{-10pt}
\end{figure*}

(1) Selected users are with high possibility to rate $i_{new}$. This can be modeled as a classification task. To achieve this, we first transform the user-item rating matrix $\mathbf{R}$ to a 0-1 matrix $\mathbf{R}^{01}$ \cite{koren:2008factorization}, where all entries with ratings are assigned to 1, and all entries with no ratings are assigned to 0 (see Figure \ref{fig:chooseToRate} (a)). Then we use FM to model them \cite{deldjoo:2016using}, where all entries equal to 1 are regarded as positive instances, and a same number of negative instances are selected from the entries equal to 0. Features contain users and attributes (without items). Labels are 1 for positive instances and 0 for negative instances. The classification model is trained based on $\mathbf{R}^{01}$ and $\mathbf{T}$. The general process is shown in Figure \ref{fig:chooseToRate} (b).

Finally, a vector $\mathbf{p}$ is defined as follows:
\begin{equation}\label{eq:high pro to rate}
\begin{aligned}
p(m) = willing\_score(u_m,i_{new}), u_m\in U,
\end{aligned}
\end{equation}
where $willing\_score(u_m,i_{new})$ is the possibility that user $u_m$ will rate item $i_{new}$, which is predicted by our learned classification model. We tend to select $u_m$ if $p(m)$ is large.

(2) Selected users' \emph{potential ratings} are diverse. Potential ratings are users' ratings on $i_{new}$ purely estimated according to $i_{new}$'s attributes (without feedback since there is no feedback yet). We expect selected users' potential ratings are diverse, so that: 1) selected users tend to have different interest. Ratings of these users would provide more information compared to ratings of similar users, and 2) the final prediction model trained on these users' feedback would not be biased to a fixed region of ratings. To choose users with diverse potential ratings, we firstly train a regression model based on $\mathbf{R}$ and $\mathbf{T}$, as shown in Figure \ref{fig:chooseToRate} (c). In this regression model, features contain users and attributes (without items). Labels are users' ratings. Once the regression model is learned, all users' potential ratings on the new item $P_r(u_m, i_{new}), u_m\in U$ can be estimated. Secondly, pair-wise diverse values among all these ratings are calculated to form the diverse matrix $\mathbf{D}$. The diverse value between potential ratings of $u_m$ and $u_n$ is defined as:
\begin{equation}\label{eq:diverse function}
\begin{aligned}
D(m,n) &= |P_r(u_m, i_{new}) - P_r(u_n, i_{new})|^{\frac{1}{2}}.
\end{aligned}
\end{equation}
Calculating $\mathbf{D}$ is  computationally expensive. However, the calculations of diverse values are independent with each other. Therefore, when applied to real world recommender systems, they can be performed parallelly with acceleration techniques such as GPU acceleration \cite{tarditi:2006accelerator}, distributed computing \cite{bokhari:2012assignment}, etc. We tend to select $u_m$ and $u_n$ together if $D(m,n)$ is large.

(3) Selected users' generated ratings are \emph{objective}. A rating on an item is objective means that this rating approximates the average of all ratings on this item, which is a good estimation of this item's quality \cite{duan:2008online, chevalier:2006effect, dellarocas:2004exploring}. We favor selecting users who always generate objective ratings in the past. Then they are expected to also generate objective ratings for $i_{new}$. We form a vector $\mathbf{o}$, which consists of all users' objective values. The objective value of user $u_m$ is defined as:
\begin{equation}\label{eq:high objective}
\begin{aligned}
o(m) &= \frac{1}{\log |I(u_m)| + 1}\cdot \frac{1}{|I(u_m)|}\cdot \sum_{i_n\in I(u_m)} (R(m,n)-\overline{R(n)})^2,
\end{aligned}
\end{equation}
where $I(u_m)$ is the item set that $u_m$ has rated. $R(m,n)$ is $u_m$'s rating on $i_n$. $\overline{R(n)}$ is the mean rating on $i_n$. $\frac{1}{\log |I(u_m)|}$ is a penalty for users who have rated few items, since a user may generate a rating that approximates $\overline{R(n)}$ by coincidence. Note that a smaller $o(m)$ indicates that $u_m$ is more objective, so we tend to select $u_m$ if $o(m)$ is small.

Users selected with this criterion would give ratings that can better reflect the quality of items, i.e. higher for items with better quality and verse vice. Therefore, with this criterion, the re-trained prediction model would generate overall higher/lower prediction ratings for new items with better/worse quality, which is more reasonable. This criterion is a complement to Criterion (2). Criterion (2) encourages the feedback ratings of selected users to have a large variance, thus it could increase the model's differentiation power in terms of different users. With Criterion (3), we want the average of feedback ratings to be higher/lower for items with better/worse quality, which could increase the model's differentiation power in terms of different new items.



(4) Selected users are \emph{representative}. A selected user is representative means that this user is similar to unselected users. Selected users should be representative so that from their feedback, we can learn more about the preference of unselected users. To achieve this, we firstly construct a similarity matrix $\mathbf{S}$ from users' rating history. That is, each user is represented as a row vector of the user-item matrix $\mathbf{R}$, then the similarity between two users can be measured based on their vectors. $\mathbf{S}$ is defined as:
\begin{equation}\label{eq:representative}
\begin{aligned}
   S(m,n)=
   \begin{cases}
   Sim(R(m,:),R(n,:))& \mbox{if $m\neq n$}\\
   0  &\mbox{if $m = n$}
   \end{cases},
\end{aligned}
\end{equation}
where $R(m,:)$ and $R(n,:)$ are vectors of users $u_m$ and $u_n$, and $Sim(R(m,:),R(n,:))$ is their similarity. In this paper, cosine similarity is used to measure it. Acceleration techniques described in Criterion (2) can also be applied to calculating $\mathbf{S}$. We tend to select one of $u_m$ and $u_n$ if $S(m,n)$ is large. Criterion (2) and Criterion (4) are highly related to the \emph{avoiding redundancy} and \emph{ensuring representativeness} described in \cite{chattopadhyay:2012batch}, which can be interpreted from the perspective of minimizing the distribution difference between the labeled and unlabeled data.

We now formulate the user selection task as an explicit mathematical optimization problem, where the objective is to select a batch of users based on above criteria. Specifically, we define a binary vector $\mathbf{q}$ with $|U|$ entries ($\mathbf{q} \in \{0, 1\}^{|U|\times 1}$), where each entry $q(m)$ denotes whether $u_m$ will be included in the batch ($q(m) = 1$) or not ($q(m) = 0$). Thus our user selection strategy (with given batch size $k$) can be expressed as the following integer quadratic programming (IQP) problem:

\begin{equation}\label{eq:initial objective function}
\begin{aligned}
   \max_\mathbf{q} \mbox{~} &{\alpha\sum_{m=1}^{|U|} q(m)p(m) + \beta \sum_{m=1}^{|U|} \sum_{n=1}^{|U|} q(m)q(n)D(m,n)}\\
   & - \gamma \sum_{m=1}^{|U|} q(m)o(m)+ \sigma \sum_{m=1}^{|U|} \sum_{n=1}^{|U|} q(m)(1-q(n))S(m,n)\\
   &\mbox{s.t.~~~} q(m)\in \{0,1\}, \forall m\mbox{~~~and~~~} \sum_{m=1}^{|U|}q(m) = k.
\end{aligned}
\end{equation}
The first term is to satisfy Criterion (1). Supposing $u_m$ is with high possibility to rate $i_{new}$ ($p(m)$ is large), then in order to optimize the objective function, $u_m$ is encouraged to be selected ($q(m)$ is encouraged to be 1). The second term is to satisfy Criterion (2). Supposing potential ratings of $u_m$ and $u_n$ are very diverse ($D(m,n)$ is large), then $u_m$ and $u_n$ are encouraged to be selected together ($q(m)$ and $q(n)$ are encouraged to be 1 together) in order to optimize the objective function. The third term is to satisfy Criterion (3), whose analysis is similar to the analysis for the first term, except that we minus this term, since we want to select $u_m$ when $o(m)$ is small. The last term enforces selected users to be similar to unselected users, ensuring representativeness, which satisfies Criterion (4). This can be similarly analyzed as for the second term. $\alpha$, $\beta$, $\gamma$ and $\sigma$ are trade-off parameters.

Equation (\ref{eq:initial objective function}) can be reformulated as:
\begin{equation}\label{eq:initial objective function 2}
\begin{aligned}
   &\alpha\mathbf{q}^T \mathbf{p} +  \beta \mathbf{q}^T \mathbf{D} \mathbf{q} - \gamma\mathbf{q}^T \mathbf{o} + \sigma \mathbf{q}^T \mathbf{S} (\mathbf{1}-\mathbf{q})\\
   =&\alpha\mathbf{q}^T \mathbf{p} +  \beta \mathbf{q}^T \mathbf{D} \mathbf{q} - \gamma\mathbf{q}^T \mathbf{o} + \sigma \mathbf{q}^T \mathbf{S}\mathbf{1}-\sigma \mathbf{q}^T \mathbf{S}\mathbf{q}\\
   =&\mathbf{q}^T(\alpha\mathbf{p} - \gamma\mathbf{o}+\sigma\mathbf{S}\mathbf{1}) + \mathbf{q}^T (\beta \mathbf{D} - \sigma\mathbf{S})\mathbf{q}\\
   =&\mathbf{q}^T diag(\alpha\mathbf{p} - \gamma\mathbf{o}+\sigma\mathbf{S}\mathbf{1})\mathbf{q} + \mathbf{q}^T (\beta \mathbf{D} - \sigma\mathbf{S})\mathbf{q}\\
    =&\mathbf{q}^T \mathbf{M} \mathbf{q}\\
   &\mbox{s.t.~~~} q(m)\in \{0,1\}, \forall m\mbox{~~~and~~~} \sum_{m=1}^{|U|}q(m) = k,
\end{aligned}
\end{equation}
where $\mathbf{1}$ is a vector with all entries equal to $1$ and $diag(\alpha\mathbf{p} - \gamma\mathbf{o}+\sigma\mathbf{S}\mathbf{1})$ is a diagonal matrix, whose $(i,i)$-th entry is equal to the $i$-th entry of ($\alpha\mathbf{p} - \gamma\mathbf{o}+\sigma\mathbf{S}\mathbf{1}$). Since we have the constraint $q(m)\in \{0,1\}, \forall m$, thus we derive $\mathbf{q}^T(\alpha\mathbf{p} - \gamma\mathbf{o}+\sigma\mathbf{S}\mathbf{1}) = \mathbf{q}^T diag(\alpha\mathbf{p} - \gamma\mathbf{o}+\sigma\mathbf{S}\mathbf{1})\mathbf{q}$. $\mathbf{M}$ is defined as:
\begin{equation}\label{eq:final matrix}
\begin{aligned}
   M(m,n)=
   \begin{cases}
   \beta D(m,n)-\sigma S(m,n)& \mbox{if $m\neq n$}\\
   \alpha p(m) - \gamma o(m) + \sigma \mathbf{S} \mathbf{1}(m)  &\mbox{if $m = n$}
   \end{cases}.
\end{aligned}
\end{equation}
Finally, the objective function is transformed to:
\begin{equation}\label{eq:final objective function}
\begin{aligned}
&\max_\mathbf{q}{\mathbf{q}^T \mathbf{M} \mathbf{q}},\\
&\mbox{s.t.~~~} q(m)\in \{0,1\}, \forall m\mbox{~~~and~~~} \sum_{m=1}^{|U|}q(m) = k.
\end{aligned}
\end{equation}
Directly solving this integer quadratic programming (IQP) problem is NP-hard. However, it can be relaxed to 1) a convex quadratic programming (QP) problem by relaxing the constraint $q(m)\in \{0,1\}$ to  $q(m)\in [0,1]$ \cite{chattopadhyay:2012batch}, or 2) a convex linear programming (LP) problem of two types, one from \cite{chattopadhyay:2012batch} and the other from \cite{chakraborty:2015batchrank}. Here we briefly describe the convex LP solution from \cite{chakraborty:2015batchrank}. It is by two steps. In step 1, compute a vector $\mathbf{v}\in R^{|U|\times 1}$ containing column sums of $\mathbf{M}$ and identify the k largest entries in $\mathbf{v}$ to derive the initial solution $\mathbf{q}_0$ (replace the k largest entries of $\mathbf{v}$ with value 1 and replace the other entries with value 0, then assign it to $\mathbf{q}_0$). In step 2, it is a iterative process as shown in \textbf{Algorithm \ref{algo:iterative process}}. Starting with initial solution $\mathbf{q}_0$, we generate a sequence of solutions $\mathbf{q}_1, \mathbf{q}_2, \cdots$ until convergence. Finally, we get the solution $\mathbf{q}$, whose 1-value entries indicate the selected user set to rate $i_{new}$.

If $\mathbf{M}$ is positive semi-definite, \textbf{Algorithm \ref{algo:iterative process}} has a guaranteed monotonic convergence. If $\mathbf{M}$ is not positive semi-definite, with a positive scalar added to the diagonal elements, this algorithm can still be run on the shifted quadratic function to guarantee a monotonic convergence \cite{yuan:2013truncated}. Due to the monotonic convergence, the quality of the solution can only improve over iterations. The iterative process converges fast, thus there is only a marginal increase of the running time. Therefore, the complexity of our algorithm is O($|U|^2$), where $|U|$ is the number of users. Refer to \cite{chakraborty:2015batchrank} for a more detailed description of the complexity analysis. In real recommender systems, $\mathbf{M}$ may be too large for the memory to load. Our algorithm can still work well in this situation. For step 1, the memory only needs to load one column of $\mathbf{M}$ at a time to calculate column sums of $\mathbf{M}$. For each iteration of step 2, the memory only needs to load one row (equal to corresponding column, $\mathbf{M}$ is symmetric) of $\mathbf{M}$ and $\mathbf{q}_{t-1}$ at a time to calculate $\mathbf{M} \cdot \mathbf{q}_{t-1}$.

\begin{algorithm}[htb!]
\caption{Iterative Process for LP Solution}\label{algo:iterative process}
\begin{algorithmic}[1]
\STATE t = 1
\STATE \textbf{repeat}
\STATE \ \ \ \ Compute $\mathbf{q}_t^\prime = \mathbf{M} \cdot \mathbf{q}_{t-1}$
\STATE \ \ \ \ Replace the k largest entries of $\mathbf{q}_t^\prime$ with value 1 and replace the other entries with value 0
\STATE \ \ \ \ $\mathbf{q}_t=\mathbf{q}_t^\prime$
\STATE \ \ \ \ t = t + 1
\STATE \textbf{until} Convergence ($\mathbf{q}_{t-1}$ is equal to $\mathbf{q}_{t-2}$)
\end{algorithmic}
\end{algorithm}
\subsection{Active Learning for a Batch of Items}
As described in the introduction section, the budget of active learning is fixed for each new item in previous active learning works. In this paper, we propose a dynamic active learning budget so that the limited active learning resources can be properly distributed. We use $new\_item_1,new\_item_2,\cdots ,new\_item_l$ to denote $l$ new items. The total budget is denoted as $k_{total}$. Budget for all new items is denoted as $\mathbf{k}\in R^{{l}\times 1}$, where $k(1), k(2), \cdots ,k(l)$ are corresponding numbers of selected users for each new item. Thus we have $k_{total} = \sum_{i=1}^l k(i)$. We propose that more budget is distributed to new items with following two features.

Firstly, these items are \emph{popular}, which means many people would be willing to rate them. In the active learning phase, since popular items tend to be rated by more selected users, thus we will get more feedback ratings if we require ratings on popular items rather than requiring ratings on unpopular items. In the prediction phase, since popular items also tend to receive more ratings from unselected users, learning more about popular items, rather than unpopular ones, will influence and generate accurate predictions for more ratings. This is a problem of whether users will rate items
(described in Criterion (1)). We use the mean of all users' willing scores to measure it:
\begin{equation}\label{eq:popular new item}
\begin{aligned}
&popular(new\_item_{i})\\
&= \frac{1}{|U|}\sum_{u_m\in U} willing\_score(u_m,new\_item_{i}),\\
& i\in \{1,2,\cdots ,l\},
\end{aligned}
\end{equation}
where $willing\_score(u_m,new\_item_{i})$ is defined in section 4.1.

Secondly, these items are \emph{controversial}, which means we are uncertain about whether they will be liked or disliked by users. For items which will be obviously favored by almost all users, we already have a high confidence what ratings users tend to give to them. In contrary, it is the controversial items that we need to learn more about. This is a problem of what ratings users will give to items (described in Criterion (2)). We use the standard deviation of potential ratings to measure it:
\begin{equation}\label{eq:fcontroversial new item}
\begin{aligned}
&controversial(new\_item_{i}) \\
&= \frac{1}{|U|}\sqrt{\sum_{u_m\in U} (P_r(u_m,new\_item_{i})-\overline{P_r(new\_item_{i})})^2},\\
& i\in \{1,2,\cdots ,l\},
\end{aligned}
\end{equation}
where $P_r(u_m,new\_item_{i})$ is defined in section 4.1. $\overline{P_r(new\_item_{i})}$ is the average of potential ratings on $new\_item_{i}$. A budget score for each new item is defined as:
\begin{equation}\label{eq:budget score}
\begin{aligned}
&budget\_score(new\_item_i)= \\
&popular(new\_item_i)+\lambda\cdot controversial(new\_item_{i}),
\end{aligned}
\end{equation}
where $\lambda$ is a parameter to balance importance of two features. Finally, budget is distributed as:
\begin{equation}\label{eq:ki}
\begin{aligned}
&k(i)=  \\
&\frac{budget\_score(new\_item_i)}{\sum_{j=1}^l budget\_score(new\_item_j)}\cdot k_{total}, i\in \{1,2,\cdots ,l\}.
\end{aligned}
\end{equation}
$k(i)$ are rounded to be integers. This equation ensures that more popular and controversial items will get more budget, and meanwhile each item has the opportunity to gain some budget.

\subsection{Rating Prediction Based on Feedback}

\begin{figure*}[htb!]
\begin{center}
\subfigure[A feedback rating is obtained for the new item ]{\includegraphics[width=0.73\columnwidth]{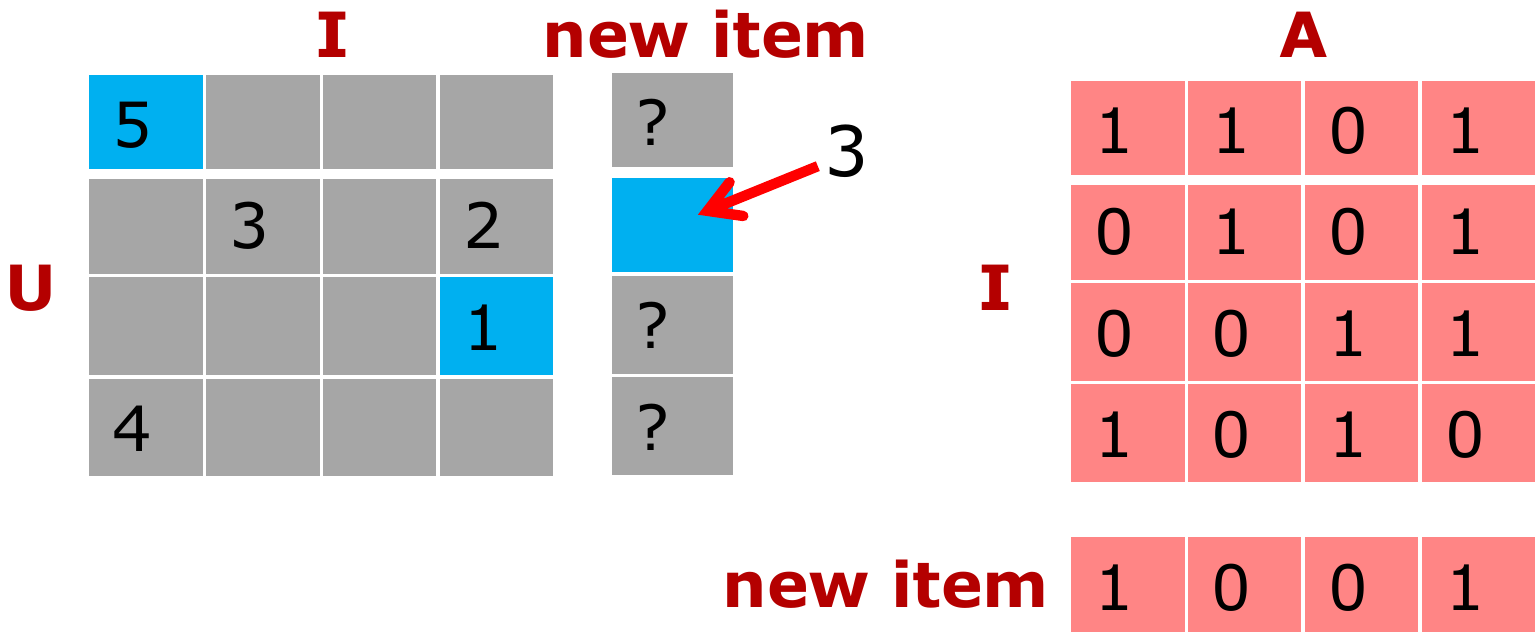}}
\subfigure[Factorization Machines for final prediction]{
\begin{minipage}[b]{0.5\textwidth}
\centering
\includegraphics[width=0.9\columnwidth]{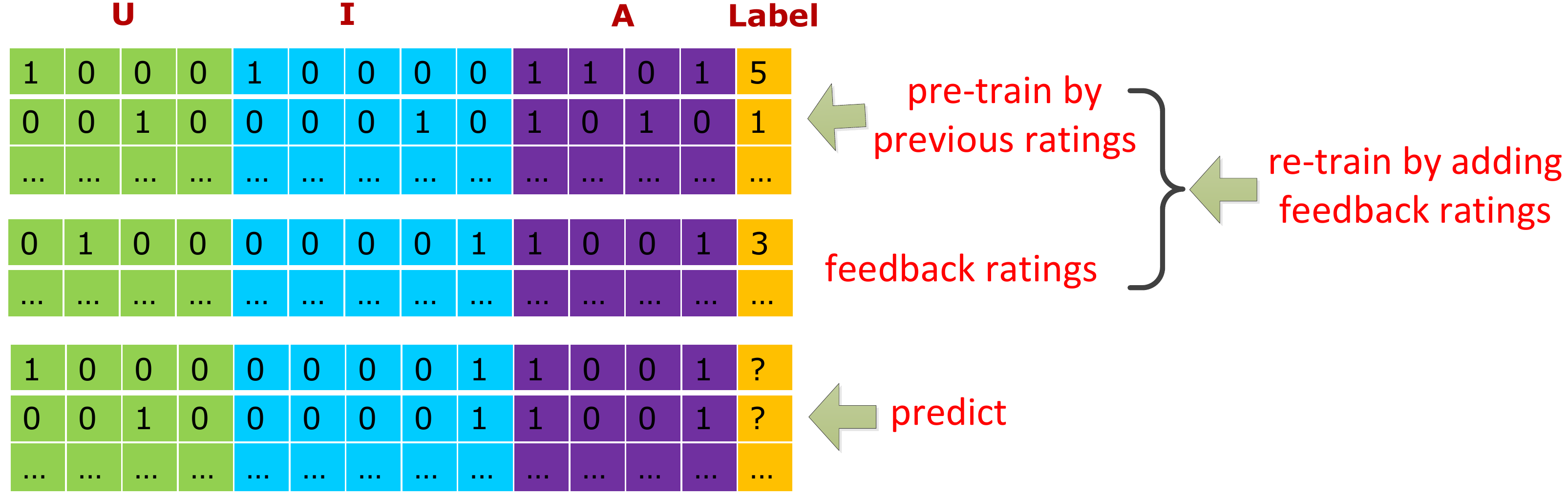}
\end{minipage}
}
\end{center}
\vspace*{-10pt}
   \caption{We firstly pre-train the regression model using previous ratings. Once feedback ratings for the new item are obtained, we re-train the model to strengthen it. (a) is to show that a feedback rating from the second user (rating `3') is obtained. (b) is to show how we re-train the model by adding users' feedback ratings. Finally, ratings of all unselected users are predicted.
}
\label{fig:final prediction}
\vspace*{-10pt}
\end{figure*}

Once selected users' feedback is obtained, we use another regression model to predict unselected users' ratings. Features in this model contain not only users and items' attributes, but also items. The instances contain both previous ratings and the newly obtained feedback ratings. Again, this is modeled by the Factorization Machines. To reduce the iteration number and accelerate the convergence speed, we firstly pre-train the regression model using previous ratings to get pre-trained parameters. Then when feedback ratings are obtained, we use these pre-trained parameters as initial parameters, all previous ratings and feedback ratings as training data, to re-train the model. Finally, ratings of all unselected users are predicted. Figure \ref{fig:final prediction} shows the detailed procedure.

The procedure of firstly pre-training and then re-training is similar to the idea of Bayesian Analysis \cite{berger:2013statistical}. That is, by pre-training on previous items, we learn users' preferences on attributes, and gain a ``prior'' understanding of users' preferences on $i_{new}$ estimated according to $i_{new}$'s attributes. Then users' feedback on $i_{new}$ enhance our understanding of $i_{new}$ and allow us to give ``posterior'' estimations for users' preferences on $i_{new}$.

\subsection{Exploitation-exploration Analysis}
As described in the introduction section, there are two goals in our task, i.e. \emph{exploitation} (exploiting ``existing knowledge'' to select users who are willing to rate new items, in order to obtain good user experience in the active learning phase) and \emph{exploration} (selecting users whose feedback can provide as much ``new knowledge'' about unselected users' preferences as possible and generate accurate rating predictions for unselected users, in order to obtain good user experience in the prediction phase). 1) The strategy of dynamic budget distributes more budget to popular items on which people are more willing to give ratings. Criterion (1) encourages users who are more willing to rate a certain new item to be selected. Thus they both contribute to improving the user experience in the active learning phase. 2) The strategy of dynamic budget and four criteria all help us to learn more about unselected users' preferences and generate more accurate rating predictions. Thus they all contribute to improving the user experience in the prediction phase. Therefore, our method considers both of these two goals (\emph{exploitation} and \emph{exploration}). In addition, we are able to adjust the parameter setting to further balance their trade-off. Specifically, once the best prediction accuracy is obtained with all parameters assigned to appropriate values, if we want to attach more importance to the user experience in the active learning phase, we just need to simply increase $\alpha$ (the weight of Criterion (1)). The reason is that, putting more weight on Criterion (1) would result in a higher rate of feedback ratings. However, increasing $\alpha$ will destroy the optimized parameter setting for rating prediction, thus the prediction accuracy would decrease.

\section{Experiments}

\subsection{Dataset}
Our proposed algorithm is evaluated on two datasets, Movielens-IMDB and Amazon. For the Movielens-IMDB dataset, ratings are collected from \emph{Movielens} \footnote{http://grouplens.org/datasets/movielens/} and attributes of movies are collected from \emph{imdbpy} \footnote{http://imdbpy.sourceforge.net/}. \emph{Ratings} \footnote{http://jmcauley.ucsd.edu/data/amazon/links.html} and \emph{attributes} \footnote{https://developer.amazon.com/} are also collected for the Amazon dataset. The statistics of these two datasets are shown in Table \ref{table:Movielens-IMDB and Amazon datasets}. In Movielens-IMDB, the number of attributes is the total number of directors, actors, genres, etc. In Amazon, the number of attributes is the total number of authors, publishers, etc. We collect ratings from the Movielens dataset rather than the \emph{official Netflix dataset} \footnote{https://www.kaggle.com/netflix-inc/netflix-prize-data}. The reason is that, items' detailed attributes are required in our setting. The attributes of movies in the Movielens dataset can be collected from \emph{http://imdbpy.sourceforge.net/} by accurately linking the movie ids in Movielens and IMDB. However, we did not figure out how to accurately obtain the attributes of movies in the Netflix dataset. For each dataset, 20\% items are randomly chosen as ``new items'' (i.e. testing items) in our conducted experiments. Ratings and attributes of the other 80\% items are used to train different models. Our goal is to generate accurate rating predictions on the testing items. Following \cite{aharon:2015excuseme, anava:2015budget}, the training-testing experiments are done once (also called \emph{holdout} \cite{tan:2006introduction}). Inspired by \cite{harpale:2008personalized}, we randomly select half of all users as the active-selection set and the remaining users form the prediction set. For all tesing items, users are selected from the active-selection set in the active learning phase. The rating prediction and the evaluation are performed for users in the prediction set.
\begin{table}[htb]\caption{Movielens-IMDB and Amazon Datasets}\label{table:Movielens-IMDB and Amazon datasets}
\vspace*{-10pt}
\centering
\begin{tabular}{|c|c|c|}
\hline
 &Movielens-IMDB&Amazon\\ \hline
Number of users&5000&973\\ \hline
Number of items&9998&5000\\ \hline
Number of ratings&5154925&97967\\ \hline
Number of attributes&255942&3840\\ \hline
\end{tabular}
\vspace*{-10pt}
\end{table}

\subsection{Compared Algorithms}
  Since in this paper, we want to handle new items with no rating, thus many previous active learning recommendation methods \cite{hofmann:2003collaborative,jin:2004bayesian,schohn:2000less}, which require at least a small amount of initial ratings, are not applicable in our task. \cite{anava:2015budget,aharon:2015excuseme} are the most related works to ours, which also address the item cold-start problem in an active learning scheme. However, the approach proposed by \cite{aharon:2015excuseme} is under an online setting in which users arrive and are decided to be given rating requests one by one and apparently it is not applicable for our task. \cite{anava:2015budget} assumes that selected users will always rate the new item (the rate of feedback ratings is 100\%), while our task is under a more realistic setting that only a subset of selected users will give feedback ratings. Thus it is unfair to compare the method in \cite{anava:2015budget} with our method and other baselines. We denote our method without dynamic budget as FMFC (Factorization Machines with Four Criteria) and our method with dynamic budget as FMFC-DB. We try our best to adapt following baselines from previous literature to compare with our proposed methods. HBRNN, LCE and FM are hybrid methods which combine both content and collaborative information. The remaining algorithms all exploit active learning to perform recommendations. The pre-train schedule in Figure \ref{fig:final prediction} is the same for all active learning methods, but the re-train schedule differs since different active learning methods have different feedback ratings.

\noindent \textbf{\emph{Hybrid-based Recommendation with Nearest Neighbor (HBRNN)}:} This method \cite{iaquinta:2007hybrid} is a combination of content-based recommendation and item-based collaborative filtering. The similarity between two items $i_m$,$i_n$ (including training items and testing items) is defined as follows:
\begin{equation}\label{eq:similarity matrix}
\begin{aligned}
sim(i_m,i_n) = cos(T(m,:),T(n,:)).
\end{aligned}
\end{equation}
where $T(m,:)$ and $T(n,:)$ are row vectors of the item-attribute matrix $\mathbf{T}$, representing $i_m$ and $i_n$ based on attributes. Once similarities between items are obtained, all users' ratings on the new item $i_{new}$ are predicted using an item-based collaborative filtering idea:
\begin{equation}\label{eq:HBRNN}
\begin{aligned}
Rating(u_m,i_{new}) = &\frac{\sum_{i_n\in I(u_m)}R(u_m,i_n) sim(i_n,i_{new})}{\sum_{i_n\in I(u_m)} sim(i_n,i_{new})}, \\
&u_m\in U,
\end{aligned}
\end{equation}
where $I(u_m)$ is the item set that $u_m$ rates. $R(u_m,i_n)$ is the rating that $u_m$ gives to $i_n$.

\noindent \textbf{\emph{Local Collective Embeddings (LCE)}:} This method \cite{saveski:2014item} also combines content-based recommendation and collaborative filtering. Different from HBRNN, which is a hybrid-based recommendation method from the nearest neighbor perspective. LCE is a hybrid-based recommendation method from the perspective of matrix factorization. In addition, it exploits the manifold structure of the data to improve the performance. We use the publicly available Matlab implementation \footnote{https://github.com/msaveski/LCE} of the LCE algorithm. Parameters are set and tuned as recommended in \cite{saveski:2014item}.

\noindent \textbf{\emph{Factorization Machines without Active Learning phase (FM)}:} This method uses Factorization Machines \cite{rendle:2012factorization} to model user behaviours. We directly use the pre-trained model in Figure \ref{fig:final prediction} to predict users' ratings on the new item $i_{new}$.

\noindent \textbf{\emph{Factorization Machines with Random Sampling in the Active Learning phase (FMRSAL)}:} In this baseline, for the new item $i_{new}$, $k$ users are randomly selected from the active-selection set for rating requests. Since these users are randomly selected and ratings are sparse in our dataset, thus the rate of feedback ratings is expected to be low. The performance improvement may be limited when compared to FM. However, rating requests are given to users without bias to any type of users, thus no one is always selected for rating requests in this user selection strategy.

\noindent \textbf{\emph{Factorization Machines with $\epsilon$-Greedy in the Active Learning phase (FM$\epsilon$GAL)}:} The $\epsilon$-Greedy algorithm is from the study of the multi-armed bandit problem \cite{feldman:2015recommendations}. We adapt it to our task as follows. For the new item $i_{new}$, we select $k$ users by $k$ sequential actions. For each action, we select the user who has the highest possibility to rate $i_{new}$ (i.e. $u_m$ with the largest $p(m)$) with probability $1-\epsilon$, and otherwise randomly select other users. When one user is selected in one action, he/she does not participate in following actions. In this user selection strategy, one more parameter, i.e. $\epsilon$, needs to be tuned. When we set $\epsilon = 0$, it is equal to our FMFC with only Criterion (1). When we set $\epsilon = 1$, it is transformed to FMRSAL. When we set $0<\epsilon < 1$, due to the randomness, rating requests are distributed to a wide range of users. Meanwhile, it can ensure a rate of feedback ratings higher than FMRSAL. In our experiments, we find no matter how $\epsilon$ varies, the performance in terms of all metrics is always between the performance of FMFC with only Criterion (1) and FMRSAL, thus we only show the experiment results with $\epsilon$ equal to 0.5 for simplicity.

\noindent \textbf{\emph{Factorization Machines with Poplar Sampling in the Active Learning phase (FMPSAL)}:} Inspired by \cite{golbandi:2011adaptive,rubens:2009output}, for the new item $i_{new}$, $k$ users who have given the most ratings to the training items are selected for rating requests. Since these users are ``frequently" rating users, they also tend to rate $i_{new}$, which can ensure a high rate of feedback ratings. Note that different from our Criterion (1), which is ``personalized" for different new items, users selected in this strategy are always the same.

\noindent \textbf{\emph{Factorization Machines with Coverage Sampling in the Active Learning phase (FMCSAL)}:} Inspired by \cite{golbandi:2010bootstrapping,rubens:2009output}, for the new item $i_{new}$, $k$ users who have highly co-rated items with other users are selected for rating requests. We define $Coverage(u_i) =\sum_j n_{ij}$, where $n_{ij}$ is the number of items that are rated by both users $u_i$ and $u_j$. Users with high Coverage values are then selected. The heuristic used by this strategy is that users co-rate the same items with many other users can better reflect other users' interest, and thus their rating behaviors are more helpful for predicting rating behaviors of other users.

\noindent \textbf{\emph{Factorization Machines with Exploration Sampling in the Active Learning phase (FMESAL)}:} As described in \cite{rubens:2015active}, exploration is important for recommendation, especially for new items as in our task. Inspired by studies about exploration in \cite{rubens:2007influence, chattopadhyay:2012batch}, for the new item $i_{new}$, $k$ users are selected for rating requests ensuring that selected users are representative of unselected users, and at the same time, selected users themselves are with high diversity. This can be achieved by optimizing the following objective function:
\begin{equation}\label{eq:FMES}
\begin{aligned}
&\max_\mathbf{q}{-\mathbf{q}^T \mathbf{S} \mathbf{q} +  \gamma \mathbf{q}^T \mathbf{S} (\mathbf{1}-\mathbf{q})},\\
   &\mbox{s.t.~~~} q(i)\in \{0,1\}, \forall i\mbox{~~~and~~~} \sum_{i=1}^{|U|}q(i) = k,
\end{aligned}
\end{equation}
where $\mathbf{q}$ and $\mathbf{S}$ are defined as in Equation (\ref{eq:initial objective function 2}). The first term is to ensure ``diversity" (selected users are dissimilar to each other) and the second term is for ``representative" (selected users are similar to unselected users). This integer quadratic programming (IQP) problem can be relaxed to a standard quadratic problem (QP) and be solved by applying many existing solvers.

\subsection{Evaluations}


In the active learning phase, we use following two metrics to measure the user experience of selected users.

\noindent \textbf{percentage of feedback ratings ($\bm{PFR}$)}: the ratio of users who give feedback ratings among all users who receive rating requests. It is formally defined as:
    \begin{equation}\label{eq:PFR}
    \begin{aligned}
    PFR = \frac{\mbox{Total number of feedback ratings}}{\mbox{Total number of rating requests}}.
    \end{aligned}
    \end{equation}
    Users giving feedback ratings are more likely to be willing to rate the item than those who do not give feedback ratings. A higher $PFR$ means that more selected users give feedback ratings, which indicates a better user experience.

\noindent \textbf{Average Selecting Times ($\bm{AST}$)}: average selecting times per user after a certain selection strategy is applied for all testing items. It is formally defined as:
    \begin{equation}\label{eq:AST}
    \begin{aligned}
    AST = \frac{\mbox{Total number of rating requests}}{\mbox{Total number of distinct users for rating requests}}.
    \end{aligned}
    \end{equation}
    A higher $AST$ means some users are always selected for rating requests, which will quickly annoy them and indicates a poorer user experience. For algorithms without active learning, i.e. HBRNN, LCE and FM, these two metrics are not measured.

In the prediction phase, we use \textbf{Root Mean Square Error ($\bm{RMSE}$)} and \textbf{Mean Absolute Error ($\bm{MAE}$)} to measure the user experience of unselected users. They are defined as follows:
\begin{equation}\label{eq:RMSE}
\begin{aligned}
&RMSE = \\
&\sqrt {\frac{1}{|R|}\sum_{(u_m,i_{new})\in R}(R(u_m,i_{new})-\tilde{R}(u_m,i_{new})})^2,
\end{aligned}
\end{equation}

\begin{equation}\label{eq:MAE}
\begin{aligned}
&MAE = \\
&\frac{1}{|R|}\sum_{(u_m,i_{new})\in R} |R(u_m,i_{new})-\tilde{R}(u_m,i_{new})|,
\end{aligned}
\end{equation}
$R$ are sets of (user, item) pairs that users give ratings to new items. $R(u_m,i_{new})$ is the rating that $u_m$ actually gives to $i_{new}$ and $\tilde{R}(u_m,i_{new})$ is the predicted rating.

For methods with no active learning, i.e. HBRNN, LCE and FM, models are directly trained on training items. For the remaining methods, given a new testing item, we select some users from the active-selection set in the active learning phase to see whether they have actual ratings on the testing item. If yes, we regard these actual ratings as feedback ratings. In the prediction phase, we exploit all feedback ratings to re-train the model. For all methods, $RMSE$ and $MAE$ are evaluated in the prediction set.

Apart from rating prediction, we can mimic a setting of top-$N$ recommendations as follows. Firstly, for all testing items, we select users from the active-selection set to get feedback and predict ratings for users in the prediction set (for HBRNN, LCE and FM, we directly predict users' ratings). Secondly, for each user, we select $N$ (we set $N = 10$) testing items with the largest predicted ratings (i.e. top-$N$ items) as the recommendation list. Finally, we regard new items with actual ratings larger than 3 as users' preferred items \cite{guan:2016weakly}. Performance is evaluated based on how many preferred items existing in the recommendation list, their actual ratings and their ranking positions. Following ranking metrics are used to evaluate the performance of top-$N$ recommendations.

\noindent \textbf{Precision, Recall}: $Precision$ is defined as the number of correctly recommended items (i.e. the number of preferred items existing in the recommendation list) divided by the number of all recommended items. $Recall$ is defined as the number of correctly recommended items divided by the total number of items which should be recommended (i.e. the number of preferred items). $Precision@k$ and $Recall@k$ are corresponding values at ranking position $k$. In our setting, there are $N = 10$ items in the recommendation list, while the number of preferred items is relatively large, thus the original $Recall$ is too small. We multiply it by a appropriate factor to get the modified $Recall$ \cite{zhu:2016heterogeneous}.

\noindent \textbf{Normalized Discount Cumulative Gain ($\bm{NDCG}$)}:
    $NDCG$ at position $k$ is defined as:
    \begin{equation}\label{eq:NDCG}
    NDCG@k = \frac{1}{IDCG} \times \sum_{i=1}^k \frac{2^{r_i}-1}{\log_2^{(i+1)}}
    \end{equation}
    where $r_i$ is the relevance rating of the item at position $i$. $IDCG$ is set so that the perfect ranking has a $NDCG$ value of 1. In our case, $r_i$ is set to be the actual rating for preferred items and 0 for the other items.

\subsection{Parameter Setting}

Before setting the parameters, we calibrate the four criteria first. The calibration contains following two steps.

Step 1: we normalize $\mathbf{p}$, $\mathbf{D}$, $\mathbf{o}$, $\mathbf{S}$ to be $\mathbf{p}^\prime$, $\mathbf{D}^\prime$, $\mathbf{o}^\prime$, $\mathbf{S}^\prime$ by standardization. Specifically, the normalization formula is: $p_i^\prime = \frac{p_i - \bar{\mathbf{p}}}{\sigma_{\mathbf{p}}}$, $D_{ij}^\prime = \frac{D_{ij} - \bar{\mathbf{D}}}{\sigma_{\mathbf{D}}}$, $o_i^\prime = \frac{o_i - \bar{\mathbf{o}}}{\sigma_{\mathbf{o}}}$, $S_{ij}^\prime = \frac{S_{ij} - \bar{\mathbf{S}}}{\sigma_{\mathbf{S}}}$, where $p_i^\prime$ and $p_i$ are the $i$-th entries of $\mathbf{p}^\prime$ and $\mathbf{p}$, respectively. $\bar{\mathbf{p}}$ and $\sigma_{\mathbf{p}}$ are the mean and standard deviation of all entries in $\mathbf{p}$. $D_{ij}^\prime$ and $D_{ij}$ are entries with the $i$-th row and $j$-th column in $\mathbf{D}^\prime$ and $\mathbf{D}$, respectively. $\bar{\mathbf{D}}$ and $\sigma_{\mathbf{D}}$ are the mean and standard deviation of all entries in $\mathbf{D}$. The denotations for $\mathbf{o}$ and $\mathbf{S}$ are similar to those in $\mathbf{p}$ and $\mathbf{D}$, respectively.

Step 2: we divide $\mathbf{D}^\prime$ and $\mathbf{S}^\prime$ by $|U|$ to be $\mathbf{D}_{new}$ and $\mathbf{S}_{new}$. There are $|U|$, $|U|*|U|$, $|U|$ and $|U|*|U|$ entries in $\mathbf{p}^\prime \in R^{|U|}$, $\mathbf{D}^\prime \in R^{|U|\times|U|}$, $\mathbf{o}^\prime \in R^{|U|}$ and $\mathbf{S}^\prime \in R^{|U|\times|U|}$, respectively. If we do not calibrate $\mathbf{p}^\prime$, $\mathbf{D}^\prime$, $\mathbf{o}^\prime$ and $\mathbf{S}^\prime$, then Eq. (\ref{eq:initial objective function}) will be vastly influenced by the second and fourth terms (due to much more entries in them). To prevent $\mathbf{D}^\prime$ and $\mathbf{S}^\prime$ from dominating $\mathbf{p}^\prime$ and $\mathbf{o}^\prime$ when optimizing Eq. (\ref{eq:initial objective function}), we divide $\mathbf{D}^\prime$ and $\mathbf{S}^\prime$ by $|U|$. $\alpha, \beta, \gamma$ and $\sigma$ in Eq. (\ref{eq:initial objective function}) are tuned based on $\mathbf{p}^\prime$, $\mathbf{D}_{new}$, $\mathbf{o}^\prime$, $\mathbf{S}_{new}$. Due to these two steps, the tuned $\alpha, \beta, \gamma$ and $\sigma$ could then have similar orders of magnitude.
\begin{table}[b]\caption{Performance Comparison of Active Learning and Rating Prediction on Movielens-IMDB}\label{table: performance comparision Movielens-IMDB}
\vspace*{-10pt}
\centering
\begin{tabular}{|c|c|c|c|c|c|c|}
\hline
 &$PFR(\%)$&$AST$&$RMSE$&$MAE$\\ \hline
HBRNN&x&x&0.8792&0.6738\\ \hline
LCE&x&x&0.8754&0.6712\\ \hline
FM&x&x&1.0364&0.7828\\ \hline
FMRSAL&5.17&\textbf{19.98}&0.9244&0.7320\\ \hline
FM$\epsilon$GAL($\epsilon = 0.5$)&14.24&23.62&0.8728&0.6701\\ \hline
FMPSAL&20.32&1998&0.8520&0.6562\\ \hline
FMCSAL&21.66&1998&0.8511&0.6549\\ \hline
FMESAL&6.35&1998&0.9149&0.7043\\ \hline
FMFC&23.04&163.76&0.8305&0.6388\\ \hline
FMFC-DB&\textbf{23.78}&140.03&\textbf{0.8231}&\textbf{0.6308}\\ \hline
\end{tabular}
\vspace*{-0pt}
\end{table}

\begin{table}[b]\caption{Performance Comparison of Active Learning and Rating Prediction on Amazon}\label{table: performance comparision Amazon}
\vspace*{-10pt}
\centering
\begin{tabular}{|c|c|c|c|c|c|c|}
\hline
 &$PFR(\%)$&$AST$&$RMSE$&$MAE$\\ \hline
HBRNN&x&x&0.8496&0.6565\\ \hline
LCE&x&x&0.8489&0.6564\\ \hline
FM&x&x&0.8781&0.6709\\ \hline
FMRSAL&2.09&\textbf{51.33}&0.8760&0.6619\\ \hline
FM$\epsilon$GAL($\epsilon = 0.5$)&8.04&55.87&0.8489&0.6561\\ \hline
FMPSAL&8.56&1000&0.8449&0.6533\\ \hline
FMCSAL&8.61&1000&0.8418&0.6511\\ \hline
FMESAL&7.63&1000&0.8540&0.6582\\ \hline
FMFC&13.87&82.24&0.8206&0.6360\\ \hline
FMFC-DB&\textbf{14.91}&71.02&\textbf{0.8055}&\textbf{0.6291}\\ \hline
\end{tabular}
\vspace*{-10pt}
\end{table}

\begin{table*}[t]\caption{Performance Comparison of top-$N$ recommendations on Movielens-IMDB}\label{table: performance comparision of top-$N$ recommendations Movielens-IMDB}
\vspace*{-10pt}
\centering
\begin{tabular}{|c|c|c|c|c|c|c|c|c|}
\hline
 &$Precision@5$&$Precision@10$&$Recall@5$&$Recall@10$&$NDCG@5$&$NDCG@10$\\ \hline
HBRNN&0.3554&0.2775&0.1042&0.1791&0.2806&0.3617\\ \hline
LCE&0.3593&0.2804&0.1054&0.1810&0.2799&0.3653\\ \hline
FM&0.2835&0.2151&0.0831&0.1388&0.2062&0.2884\\ \hline
FMRSAL&0.3017&0.2324&0.0885&0.1499&0.2252&0.3087\\ \hline
FM$\epsilon$GAL($\epsilon = 0.5$)&0.3601&0.2821&0.1055&0.1821&0.2801&0.3679\\ \hline
FMPSAL&0.3787&0.3025&0.1111&0.1952&0.3302&0.4114\\ \hline
FMCSAL&0.3869&0.3086&0.1134&0.1991&0.3298&0.4182\\ \hline
FMESAL&0.3208&0.2519&0.0941&0.1625&0.2497&0.3312\\ \hline
FMFC&0.4486&0.3591&0.1316&0.2317&0.3916&0.4738\\ \hline
FMFC-DB&\textbf{0.4791}&\textbf{0.3898}&\textbf{0.1405}&\textbf{0.2515}&\textbf{0.4436}&\textbf{0.5241}\\ \hline
\end{tabular}
\vspace*{-0pt}
\end{table*}

\begin{table*}[htb!]\caption{Performance Comparison of top-$N$ recommendations on Amazon}\label{table: performance of top-$N$ recommendations Amazon}
\vspace*{-10pt}
\centering
\begin{tabular}{|c|c|c|c|c|c|c|}
\hline
 &$Precision@5$&$Precision@10$&$Recall@5$&$Recall@10$&$NDCG@5$&$NDCG@10$\\ \hline
HBRNN&0.2638&0.2061&0.2162&0.3747&0.2871&0.3732\\ \hline
LCE&0.2671&0.2080&0.2189&0.3782&0.2903&0.3764\\ \hline
FM&0.2004&0.1437&0.1643&0.2613&0.2121&0.2947\\ \hline
FMRSAL&0.2195&0.1603&0.1799&0.2915&0.2324&0.3176\\ \hline
FM$\epsilon$GAL($\epsilon = 0.5$)&0.2673&0.2101&0.2191&0.3820&0.2953&0.3845\\ \hline
FMPSAL&0.2879&0.2310&0.2360&0.4200&0.3345&0.4201\\ \hline
FMCSAL&0.2933&0.2385&0.2404&0.4336&0.3404&0.4199\\ \hline
FMESAL&0.2388&0.1825&0.1957&0.3318&0.2543&0.3393\\ \hline
FMFC&0.3432&0.2859&0.2813&0.5198&0.3994&0.4805\\ \hline
FMFC-DB&\textbf{0.3818}&\textbf{0.3205}&\textbf{0.3130}&\textbf{0.5827}&\textbf{0.4517}&\textbf{0.5335}\\ \hline
\end{tabular}
\vspace*{-10pt}
\end{table*}
$k$, $\alpha$, $\beta$, $\gamma$ and $\sigma$ are the main parameters in our paper. $k$ is the number of selected users for active learning. We empirically set $k = 25$ for each testing item to tune the other parameters. There are four parameters $\alpha$, $\beta$, $\gamma$ and $\sigma$ to trade off the importance of different criteria. In fact, they can be multiplied by an arbitrary scaling factor, so there are exactly three free parameters. We fix $\alpha = 1$ and tune the other three free parameters by grid search. We use $RMSE$ as the tuning metric, where $RMSE$ is measured by cross-validation on the training data (users are also split to the active-selection set and the prediction set). For the Movielens-IMDB dataset, the final tuned parameters are $\alpha = 1, \beta = 0.3, \gamma = 0.1, \sigma = 0.1$. We regard the performance measured on all testing items using this parameter setting as the performance of FMFC. Furthermore, we fix $\alpha = 1, \beta = 0.3, \gamma = 0.1, \sigma = 0.1$, and implement FMFC-DB, i.e. our method with dynamic active learning budget. The total budget $k_{total}$ (see section 4.2) is set to be $25 \times l$, where $l$ is the number of testing items. We regard the performance measured in this setting as the performance of FMFC-DB. For the other active learning baselines, the performance is measured with the number of selected users equal to $25$ for each testing item. For the Amazon dataset, the final tuned parameters are $\alpha = 1, \beta = 0.3, \gamma = 0.03, \sigma = 0.1$. The other settings are the same as for the Movielens-IMDB dataset.

\subsection{Results and Analysis}
\subsubsection{Algorithm Comparison}

We now compare our methods with all baselines. All the performance is measured on testing items. As shown in Table \ref{table: performance comparision Movielens-IMDB} and Table \ref{table: performance comparision Amazon}, our methods (FMFC and FMFC-DB) outperform the other baselines in terms of $RMSE$ and $MAE$, which indicates our methods have the highest prediction accuracy in the prediction phase. Our methods also perform the best in the task of top-$N$ recommendations according to Table \ref{table: performance comparision of top-$N$ recommendations Movielens-IMDB} and Table \ref{table: performance of top-$N$ recommendations Amazon}. Factorization Machines with different active learning strategies perform better than Factorization Machines without active learning (FM). This is easy to understand since feedback ratings give us more understanding of testing items, which can be exploited to enhance the prediction model. As methods without active learning, HBRNN and LCE perform better than FM and even better than active learning methods FMRSAL and FMESAL. The reason may be that HBRNN and LCE can make better use of both content and collaborative information than FM. LCE has a slightly better performance than HBRNN. The reason may be that it exploits the manifold structure of the data to improve the performance of hybrid recommendations. FMPSAL, FMCSAL, FMFC and FMFC-DB perform better than the other three active learning methods. The main reason is that these four methods can ensure a high rate of feedback ratings (high $PFR$), which is the domain factor that influences the prediction accuracy (we will analyze this in the next section). Our methods outperform FMPSAL and FMCSAL because 1) our methods achieve higher rates of feedback ratings than FMPSAL and FMCSAL, and 2) our methods consider not only the rate of feedback ratings (Criterion (1)), but also other three factors (Criteria (2), (3), (4)) to improve the prediction accuracy. $PFR$ and $AST$ are both metrics that measure the user experience in the active learning phase. There is no active learning phase for HBRNN, LCE and FM, thus we compare the other methods in terms of $PFR$ and $AST$. For $PFR$, FMRSAL and FMESAL have rather few feedback ratings, which indicates they always give rating requests to users who do not really want to rate them. Thus these two methods negatively influence the user experience. FM$\epsilon$GAL has a relatively higher $PFR$. The other active learning methods all have a considerable number of feedback ratings. For $AST$, due to the natural randomness, FMRSAL undoubtedly performs the best with the lowest $AST$ and FM$\epsilon$GAL performs the second best. FMPSAL, FMCSAL and FMESAL select the same user set to rate all testing items and it will certainly annoy them. Overall, our methods are the best when considering all these metrics. When comparing FMFC with FMFC-DB, it can be seen that our proposed dynamic active learning budget can further improve the performance in terms of all metrics.

The significant test is performed to show whether the differences of our experiment results are statistically significant. Paired t-tests are conducted to compare the differences between the experiment performance of (1) our two proposed methods (i.e. FMFC and FMFC-DB), (2) FMFC and all baselines, and (3) FMFC-DB and all baselines. Specifically, for each dataset, we repeat the training-testing experiments for $100$ times (the testing item set is independently chosen in different times). Then given a certain metric, each method would generate $100$ metric values. The inputs of paired t-test are two sets of metric values, with each set corresponding to one compared method. Since we want to validate that one method is better than the other, we use one-tailed hypothesis. Statistical significance is set at $p<0.05$. The results show that, in terms of all metrics except for $AST$, (1) FMFC-DB has significantly better performance than FMFC, and (2) compared to all baselines, both FMFC and FMFC-DB have significantly better performance.

\subsubsection{Criteria Analysis}
\begin{table}[tb]\caption{Percentage of Feedback Ratings on Movielens-IMDB and Amazon}\label{table: Percentage of Feedback Ratings on Movielens-IMDB and Amazon}
\scriptsize
\vspace*{-10pt}
\centering
\begin{tabular}{|c|c|c|c|c|c|c|}
\hline
 &\multicolumn{2}{|c|}{$PFR(\%)$ on Movielens-IMDB}&\multicolumn{2}{|c|}{$PFR(\%)$ on Amazon}\\ \hline
 &FMFC&FMFC-DB&FMFC&FMFC-DB\\ \hline
No Criterion (1)&9.91&10.16&2.25&2.57\\ \hline
Original&\textbf{23.04}&\textbf{23.78}&\textbf{13.87}&\textbf{14.91}\\ \hline
\end{tabular}
\vspace*{-0pt}
\end{table}

\begin{table}[htb!]\caption{The Average Diverse Value of Selected Users' Ratings on Movielens-IMDB and Amazon}\label{table: The Average Diverse Value of Selected Users' Ratings on Movielens-IMDB and Amazon}
\scriptsize
\vspace*{-10pt}
\centering
\begin{tabular}{|c|c|c|c|c|c|c|}
\hline
 &\multicolumn{2}{|c|}{\multirow{2}{1.1in}{The Average Diverse Value on Movielens-IMDB}}&\multicolumn{2}{|c|}{\multirow{2}{1.1in}{The Average Diverse Value on Amazon}}\\
 &\multicolumn{2}{|c|}{}&\multicolumn{2}{|c|}{} \\
 \hline
 &FMFC&FMFC-DB&FMFC&FMFC-DB\\ \hline
No Criterion (2)&1.21&1.23&1.13&1.16\\ \hline
Original&\textbf{1.37}&\textbf{1.39}&\textbf{1.19}&\textbf{1.22}\\ \hline
\end{tabular}
\vspace*{-0pt}
\end{table}

\begin{table}[htb!]\caption{The Difference Between the Average Rating of Selected Users and the Average Rating of All Users on Movielens-IMDB and Amazon}\label{table: The Objective Value of Selected Users' Ratings on Movielens-IMDB and Amazon}
\scriptsize
\vspace*{-10pt}
\centering
\begin{tabular}{|c|c|c|c|c|c|c|}
\hline
 &\multicolumn{2}{|c|}{\multirow{2}{0.9in}{The Difference on Movielens-IMDB}}&\multicolumn{2}{|c|}{\multirow{2}{1.1in}{The Difference on Amazon}}\\
 &\multicolumn{2}{|c|}{}&\multicolumn{2}{|c|}{} \\
 \hline
 &FMFC&FMFC-DB&FMFC&FMFC-DB\\ \hline
\multirow{2}{0.8in}{No Criterion (3) $(|\bar{r}_{no\_c3} - \bar{r}_{all}|)$}&\multirow{2}{0.2in}{1.22}&\multirow{2}{0.2in}{1.19}&\multirow{2}{0.2in}{1.43}&\multirow{2}{0.2in}{1.39}\\
{}&{}&{}&{}&{} \\\hline
\multirow{2}{0.8in}{Original $(|\bar{r}_{ours} - \bar{r}_{all}|)$}&\multirow{2}{0.2in}{\textbf{0.97}}&\multirow{2}{0.2in}{\textbf{0.89}}&\multirow{2}{0.2in}{\textbf{1.22}}&\multirow{2}{0.2in}{\textbf{1.19}}\\
{}&{}&{}&{}&{} \\\hline
\end{tabular}
\vspace*{-0pt}
\end{table}

\begin{table}[htb!]\caption{The Average Similarity Value Between Selected and Unselected Users on Movielens-IMDB and Amazon}\label{table: The Average Similarity Value Between Selected and Unselected Users on Movielens-IMDB and Amazon}
\scriptsize
\vspace*{-10pt}
\centering
\begin{tabular}{|c|c|c|c|c|c|c|}
\hline
 &\multicolumn{2}{|c|}{\multirow{2}{1.1in}{The Average Similarity Value on Movielens-IMDB}}&\multicolumn{2}{|c|}{\multirow{2}{1.1in}{The Average Similarity Value on Amazon}}\\
 &\multicolumn{2}{|c|}{}&\multicolumn{2}{|c|}{} \\
 \hline
  &FMFC&FMFC-DB&FMFC&FMFC-DB\\ \hline
No Criterion (4)&0.54&0.56&0.46&0.49\\ \hline
Original&\textbf{0.61}&\textbf{0.67}&\textbf{0.52}&\textbf{0.58}\\ \hline
\end{tabular}
\vspace*{-0pt}
\end{table}
As mentioned in section 4.1, to generate more accurate rating predictions of the new item, our methods select users based on four criteria. In this section, we firstly validate whether each criterion works as they claim. Then we validate their contributions to the final prediction performance.\\\\
\textbf{Criterion (1): Selected users are with high possibility to rate $i_{new}$}

We remove Criterion (1) in FMFC and FMFC-DB to see how $PFR$ varies. The results are shown in Table \ref{table: Percentage of Feedback Ratings on Movielens-IMDB and Amazon}. Without Criterion (1), the $PFR$ decreases dramatically for both FMFC and FMFC-DB. Since a higher $PFR$ means that more selected users rate $i_{new}$, the result validates the effectiveness of Criterion (1).\\

\noindent\textbf{Criterion (2): Selected users' potential ratings are diverse}

The purpose of selecting users with diverse potential ratings is to ensure these users' actual ratings also tend to be diverse, so that the final prediction model is not biased to a fixed region of ratings. We remove Criterion (2) in FMFC and FMFC-DB to see how the average diverse value (defined in Eq. (\ref{eq:diverse function})) of selected users' actual ratings varies. As shown in Table \ref{table: The Average Diverse Value of Selected Users' Ratings on Movielens-IMDB and Amazon}, without Criterion (2), the average diverse value decreases for both FMFC and FMFC-DB. The result validates the effectiveness of Criterion (2).\\

\noindent\textbf{Criterion (3): Selected users' generated ratings are objective}

The insight of Criterion (3) is to let the average rating of users selected by our methods (denoted as $\bar{r}_{ours}$) approximate the average rating of all users (denoted as $\bar{r}_{all}$). We measure the difference between $\bar{r}_{ours}$ and $\bar{r}_{all}$, i.e. $|\bar{r}_{ours} - \bar{r}_{all}|$. Meanwhile, we calculate the average rating of users selected without Criterion (3) (denoted as $\bar{r}_{no\_c3}$) and measure $|\bar{r}_{no\_c3} - \bar{r}_{all}|$. The result is shown in Table \ref{table: The Objective Value of Selected Users' Ratings on Movielens-IMDB and Amazon}. We can see that $|\bar{r}_{no\_c3} - \bar{r}_{all}| > |\bar{r}_{ours} - \bar{r}_{all}|$ for both FMFC and FMFC-DB, which indicates that Criterion (3) can actually make the average rating of selected users closer to $\bar{r}_{all}$.\\

\noindent\textbf{Criterion (4): Selected users are representative}

The insight of Criterion (4) is to let the selected users similar to unselected users. We measure the average similarity value between selected and unselected users with/without Criterion (4) to validate this criterion. The result is shown in Table \ref{table: The Average Similarity Value Between Selected and Unselected Users on Movielens-IMDB and Amazon}. We can see that without Criterion (4), the average similarity value declines, which indicates that Criterion (4) can actually make the selected users more similar to unselected users.

\begin{figure}[t!]
\begin{center}
\subfigure[Movielens-IMDB]{\includegraphics[scale=0.33]{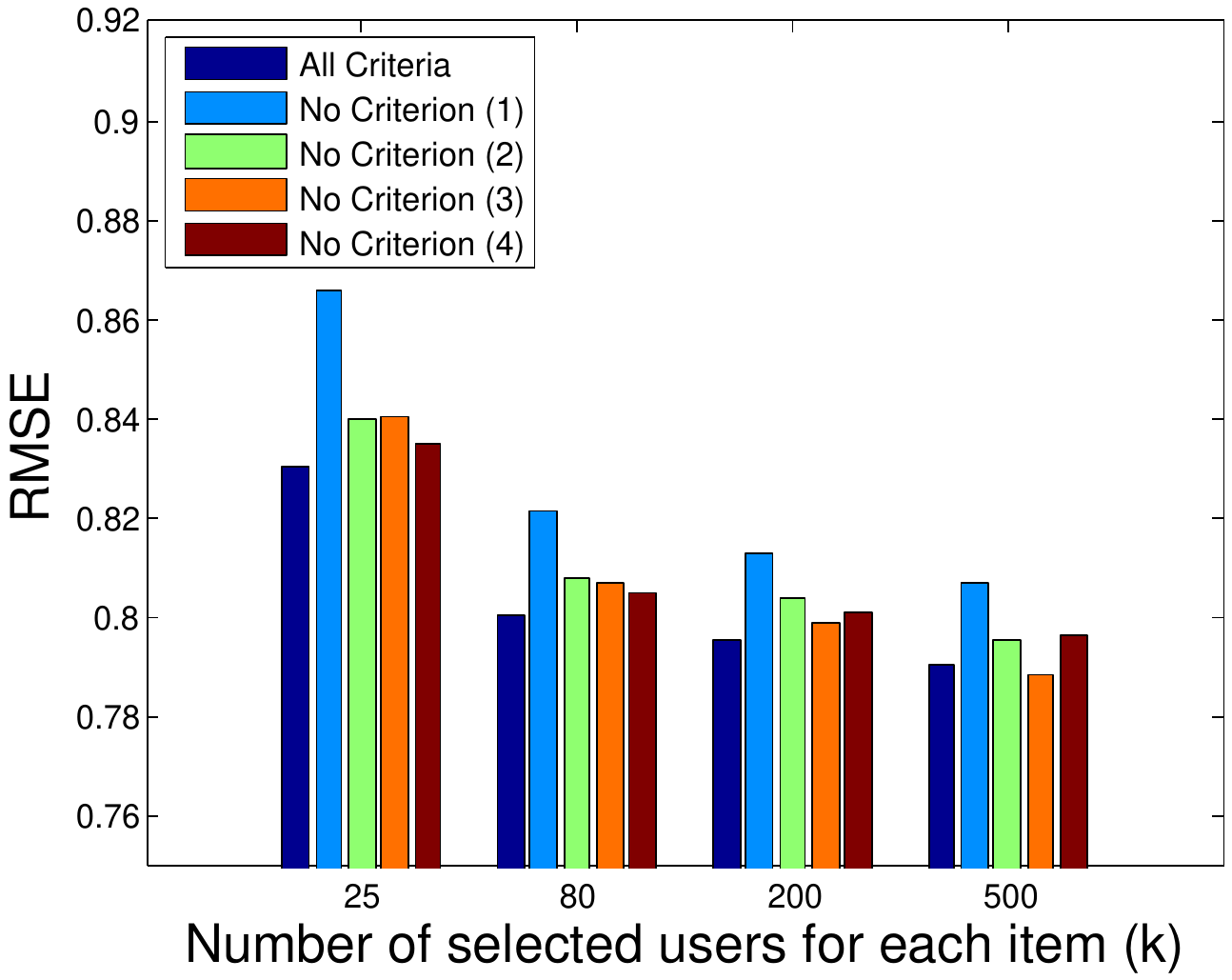}}
\subfigure[Amazon]{\includegraphics[scale=0.33]{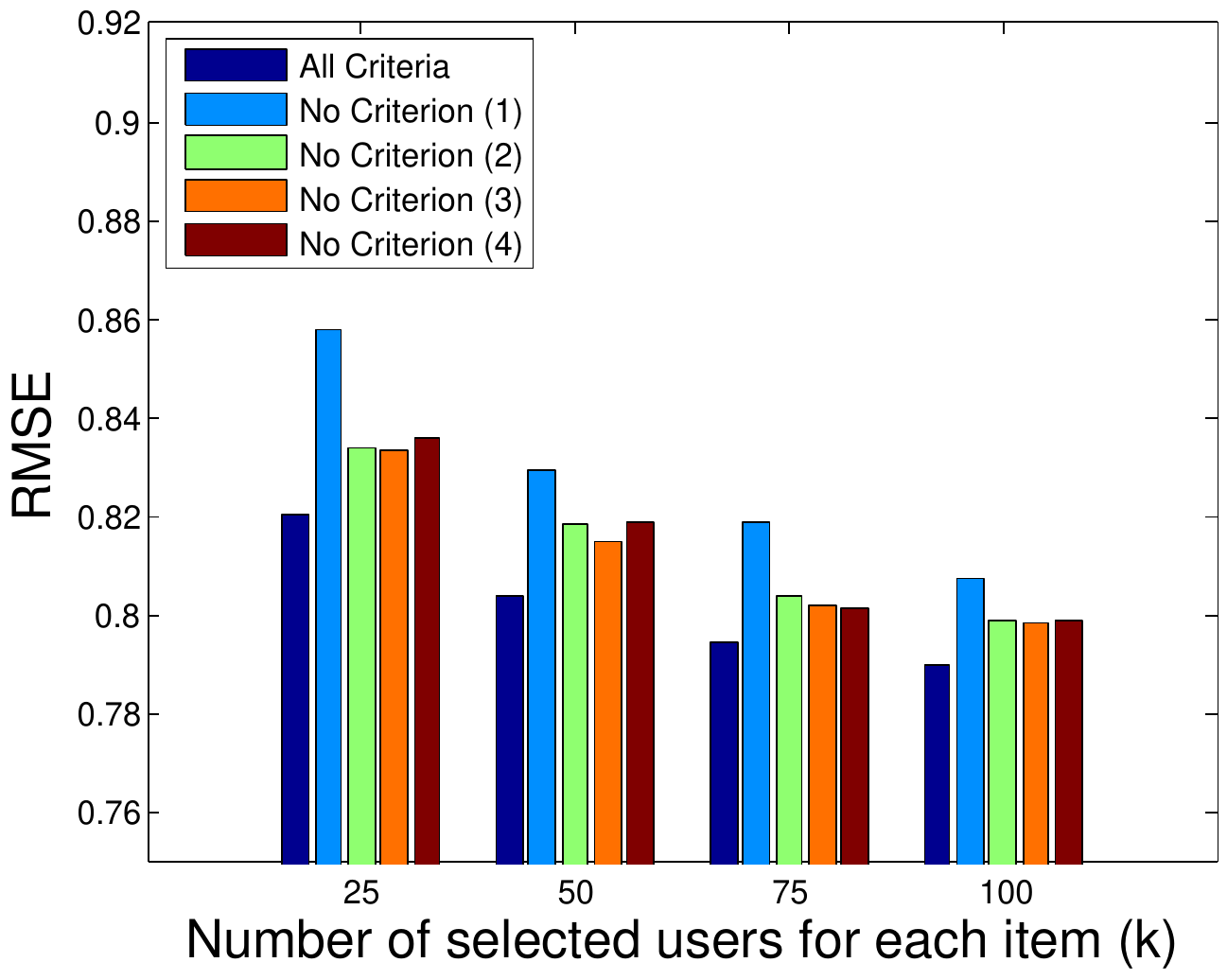}}
\end{center}
\vspace*{-10pt}
   \caption{Performance of FMFC measured by $RMSE$ when we remove four criteria one at a time and vary $k$. \emph{All Criteria} refers to the performance measured when considering all four criteria. \emph{No Criterion (1)} refers to the performance measured when we remove Criterion (1) (i.e. $\alpha = 0$). Similarly, \emph{No Criterion (2)}, \emph{No Criterion (3)} and \emph{No Criterion (4)} refer to performance measured under corresponding settings.
}
\label{fig:featurecut}
\vspace*{-10pt}
\end{figure}

We further validate the contribution of each criterion to the final prediction performance. The results are shown in Figure \ref{fig:featurecut}. $RMSE$ increases when we remove each criterion, which indicates each criterion contributes to the prediction improvement. $RMSE$ increases the most when we remove Criterion (1), which indicates this criterion is a domain factor that influences the prediction improvement. $k$ is the number of selected users. $RMSE$ decreases when $k$ increases. This is easy to understand, since larger $k$ leads to more feedback ratings, which will give us more understanding of the new item and thus generate more accurate predictions.
\subsubsection{Dynamic Budget Analysis}
\begin{figure}[t!]
\begin{center}
\subfigure[$RMSE$ on Movielens-IMDB]{\includegraphics[scale=0.3]{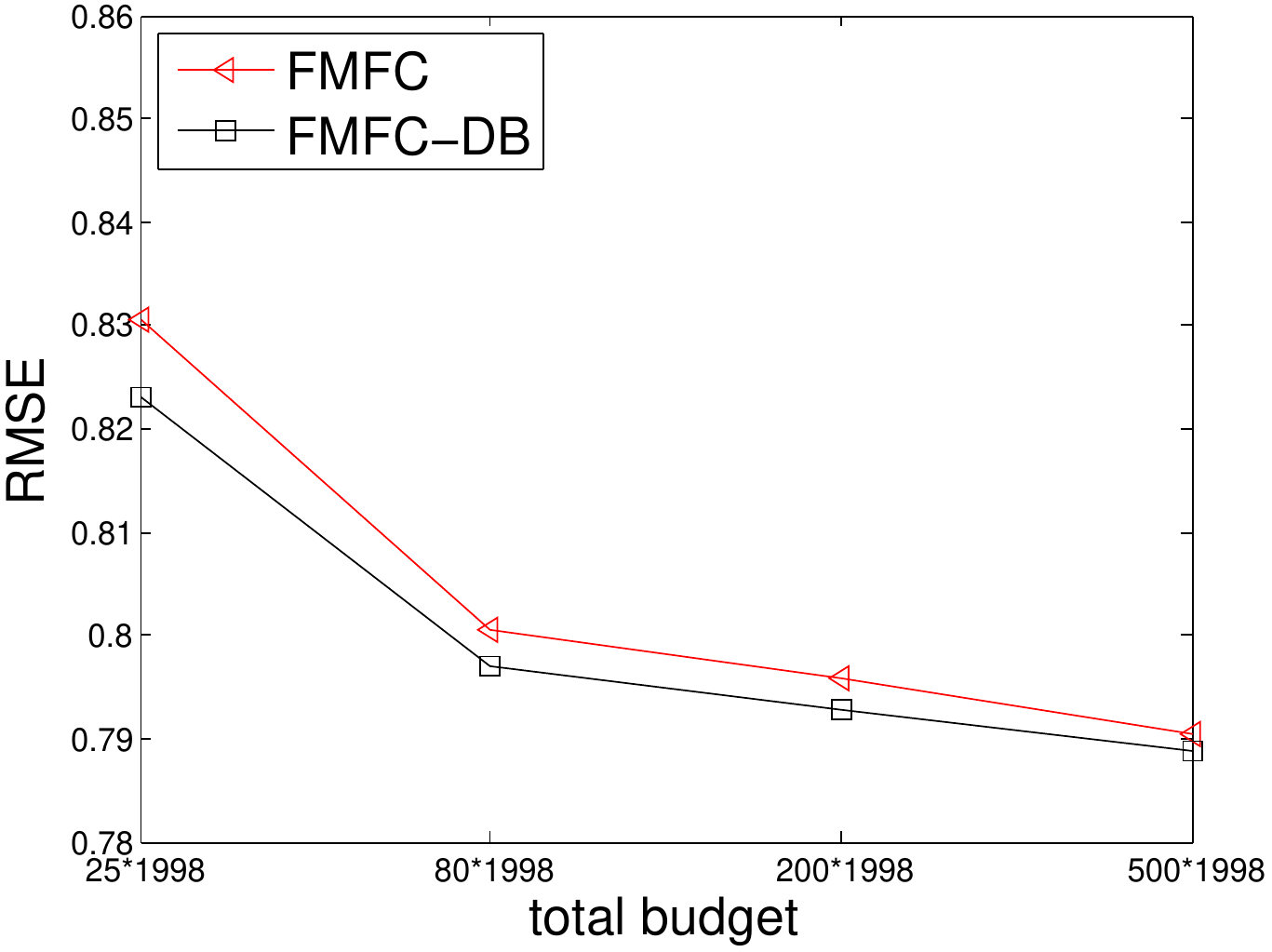}}
\subfigure[$PFR$ on Movielens-IMDB]{\includegraphics[scale=0.3]{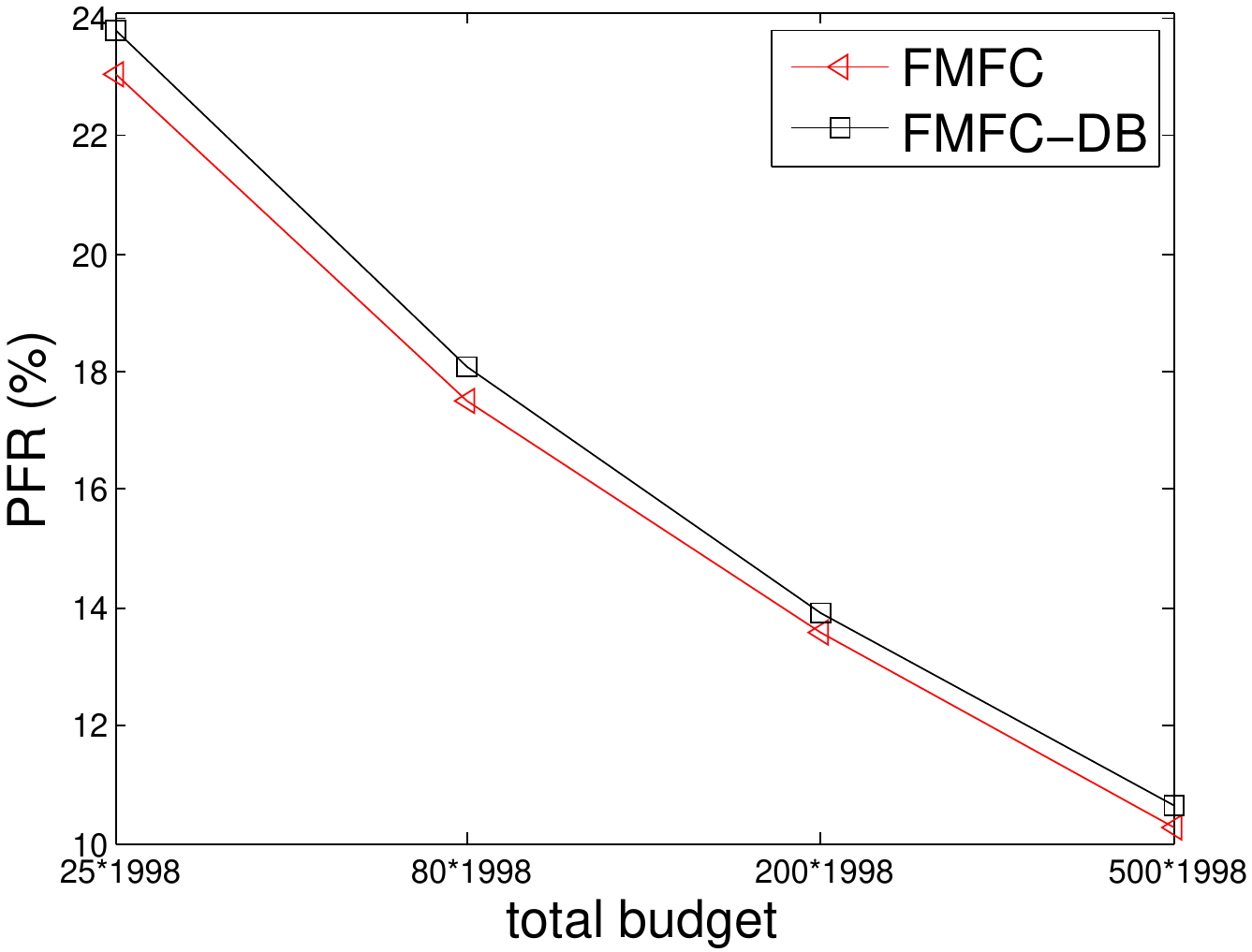}}
\subfigure[$RMSE$ on Amazon]{\includegraphics[scale=0.3]{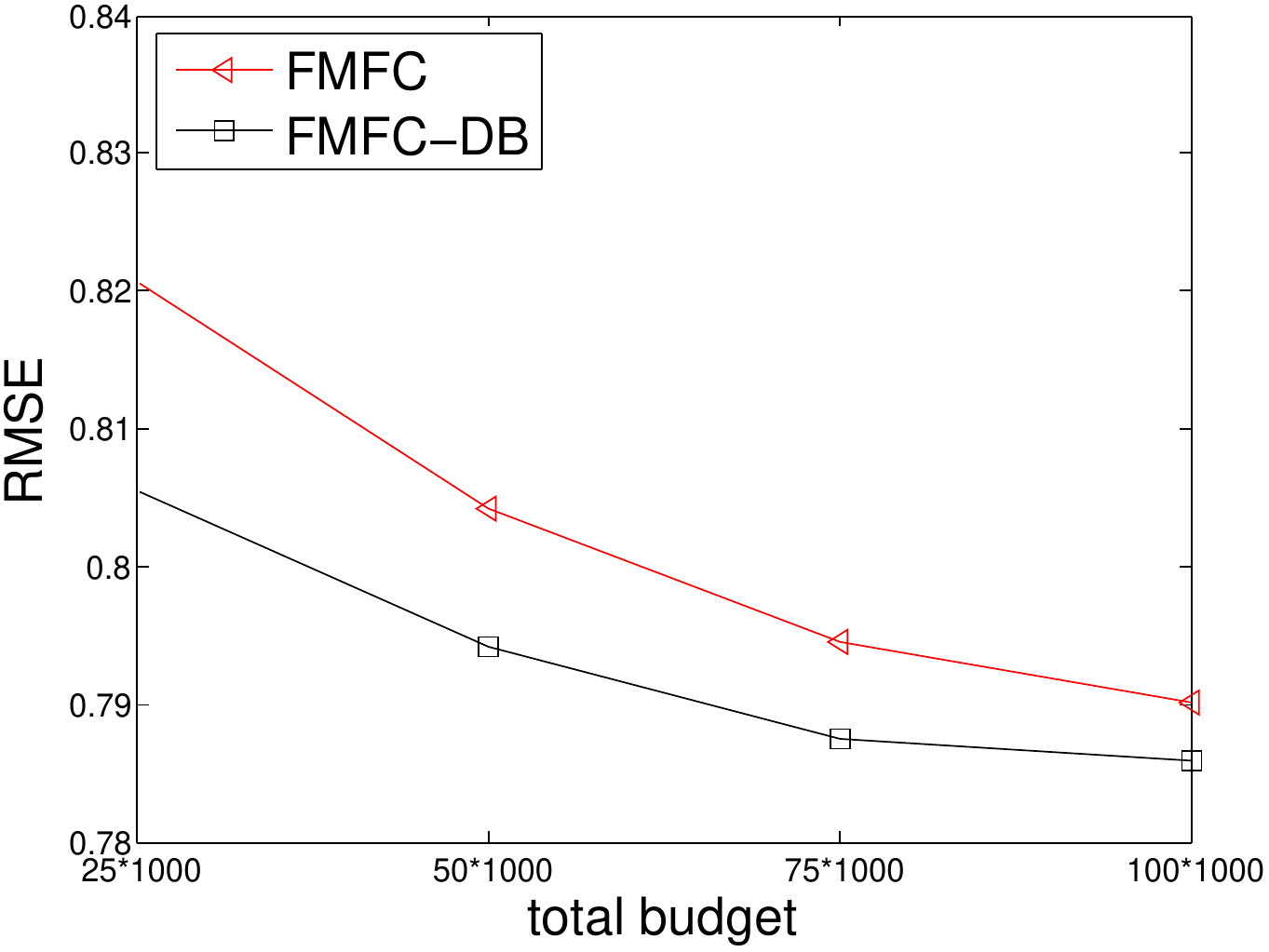}}
\subfigure[$PFR$ on Amazon]{\includegraphics[scale=0.3]{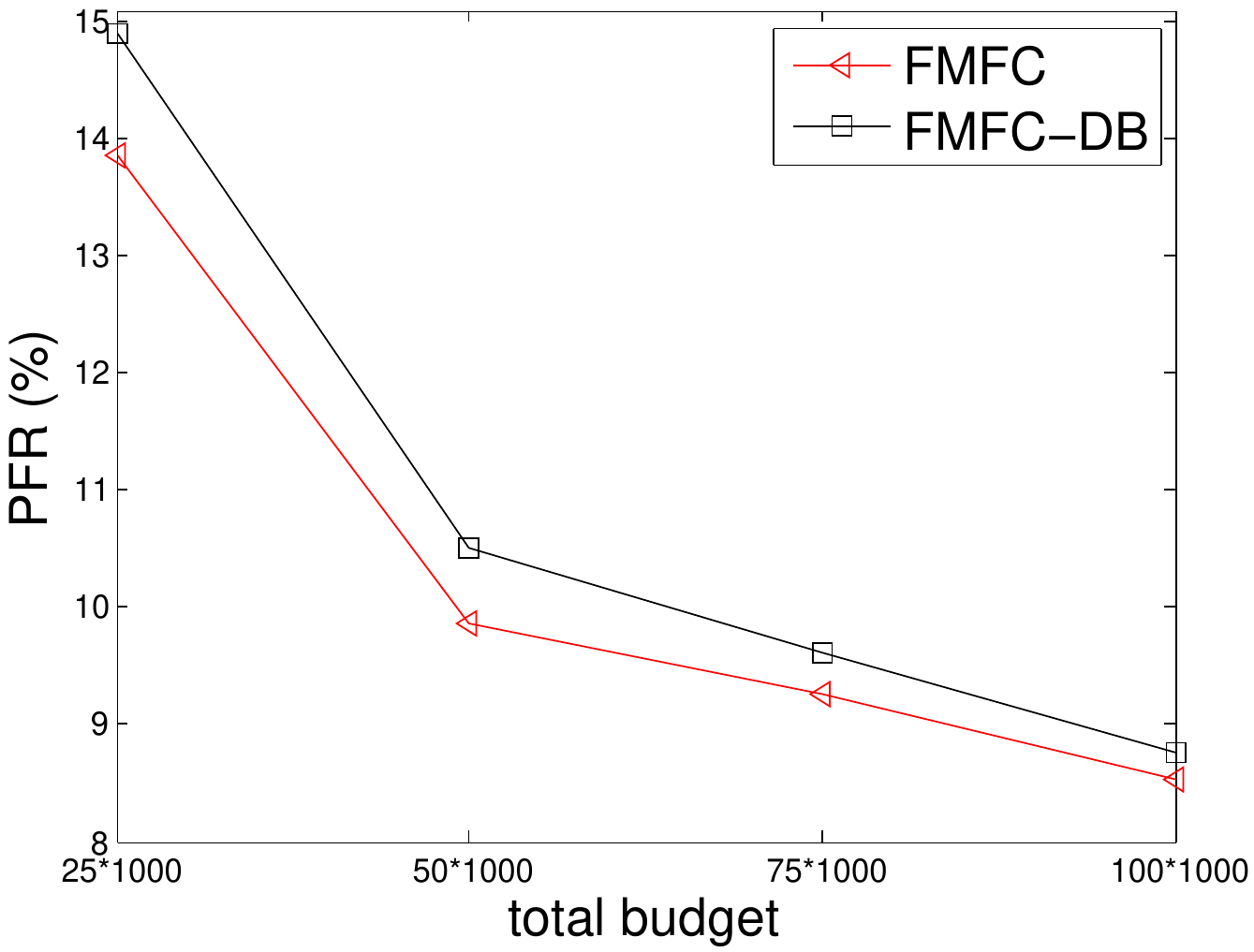}}
\end{center}
\vspace*{-10pt}
   \caption{Performance measured by $RMSE$ (for the prediction phase) and $PFR$ (for the active learning phase) when we vary the total active learning budget for both FMFC and FMFC-DB.
}
\label{fig:dynamic_budget}
\end{figure}

As mentioned in section 4.2, we use the strategy of dynamic budget to properly distribute limited active learning resources. As shown in Figure \ref{fig:dynamic_budget}, for different values of the total budget, the dynamic budget all contributes to the performance improvement in terms of both $RMSE$ and $PFR$. The improvement is narrowed when the total budget increases. This is because the dynamic budget is proposed to address the problem of limited active learning resources. When the total budget is sufficient, this strategy provides less help.
\subsubsection{Exploitation-exploration Analysis}
\begin{figure}[t!]
\begin{center}
\subfigure[$RMSE$ on Movielens-IMDB]{\includegraphics[scale=0.3]{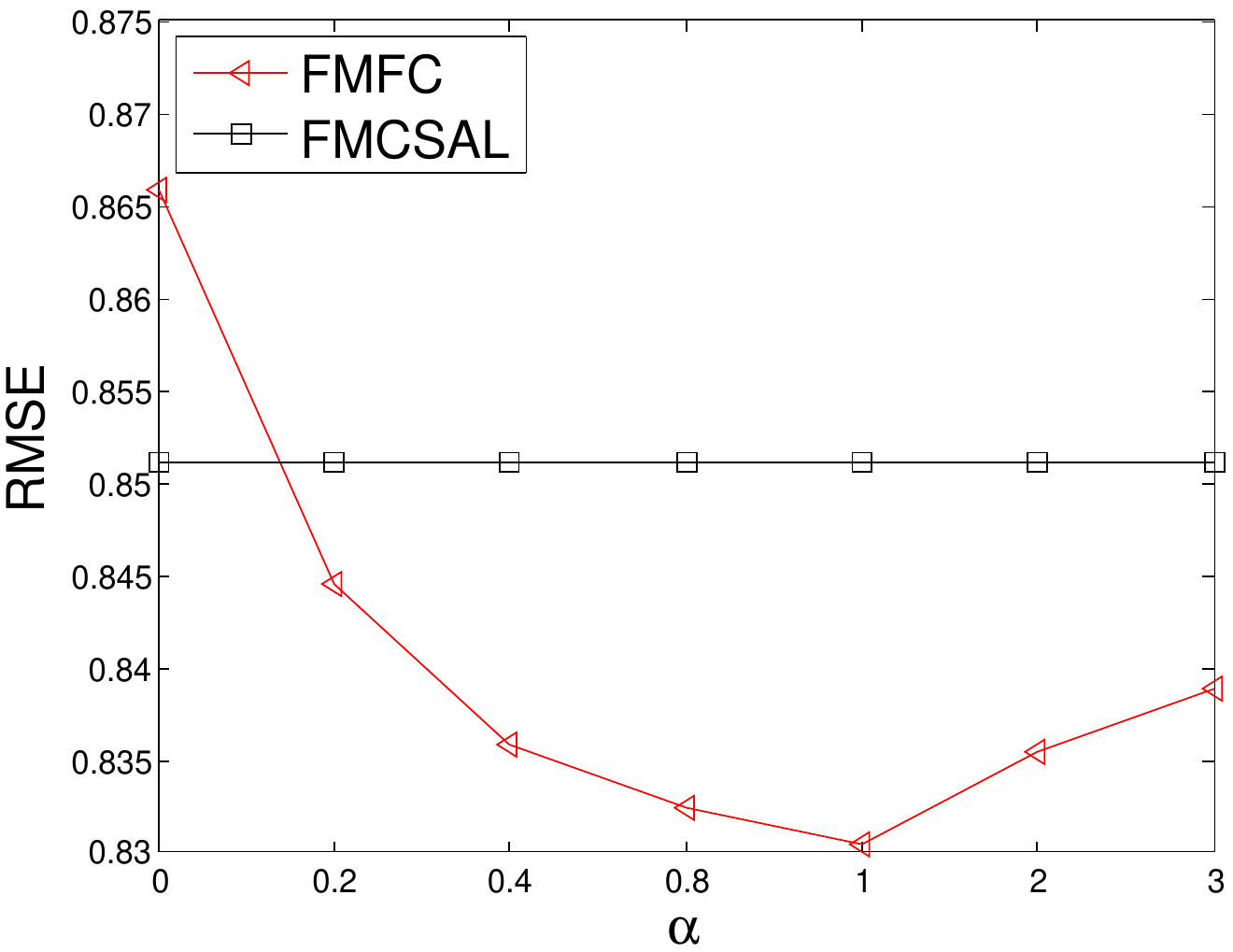}}
\subfigure[$PFR$ on Movielens-IMDB]{\includegraphics[scale=0.3]{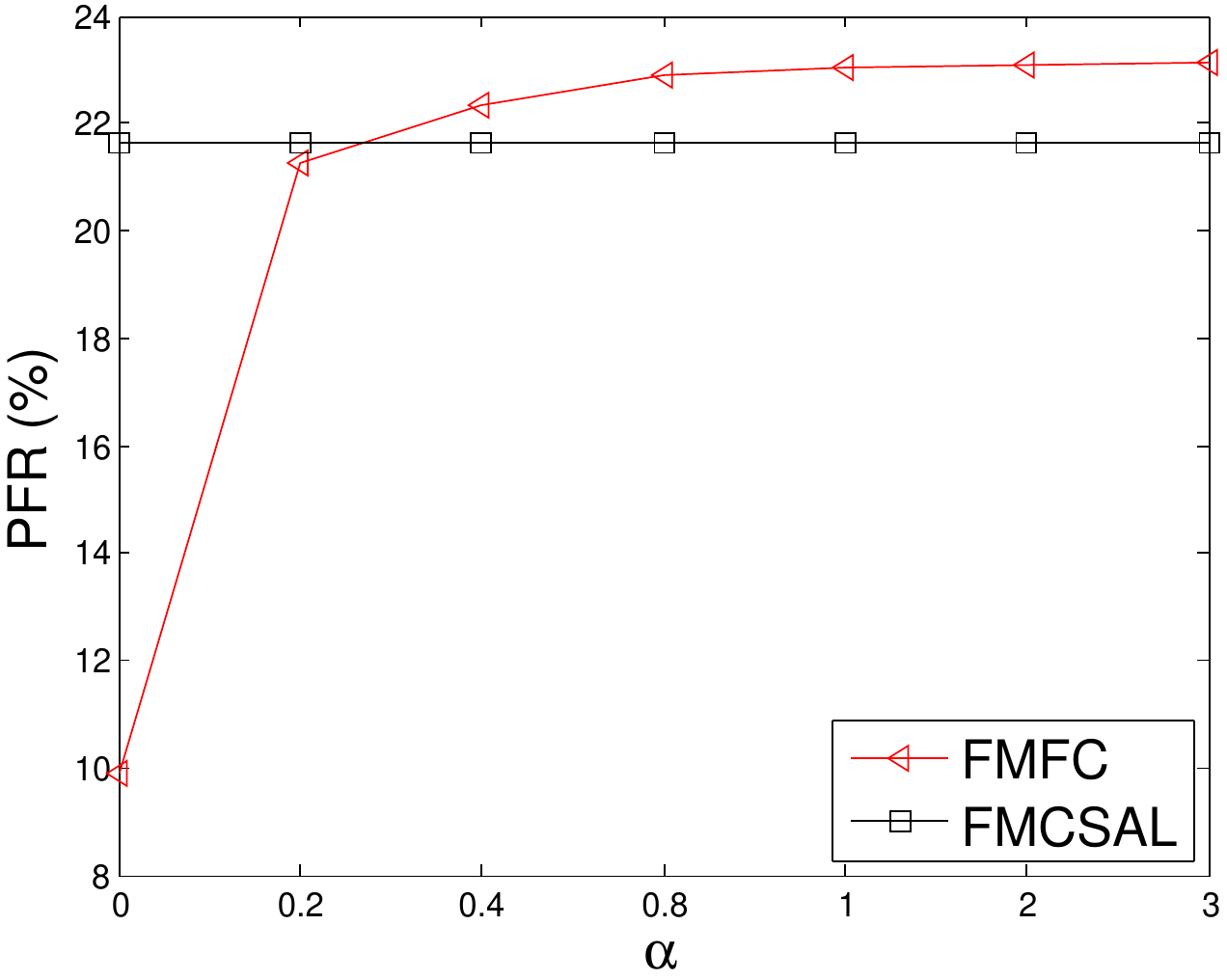}}
\subfigure[$RMSE$ on Amazon]{\includegraphics[scale=0.3]{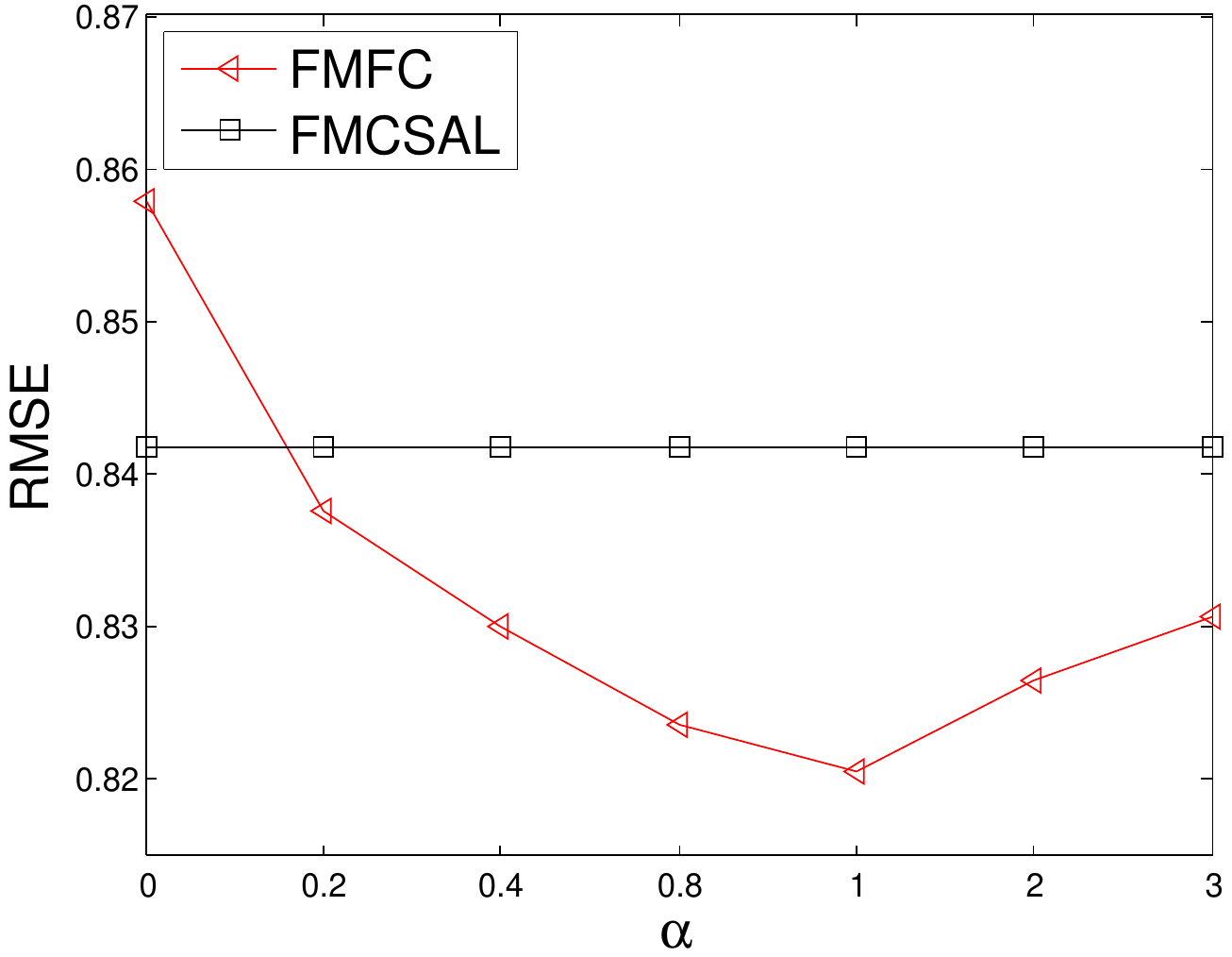}}
\subfigure[$PFR$ on Amazon]{\includegraphics[scale=0.3]{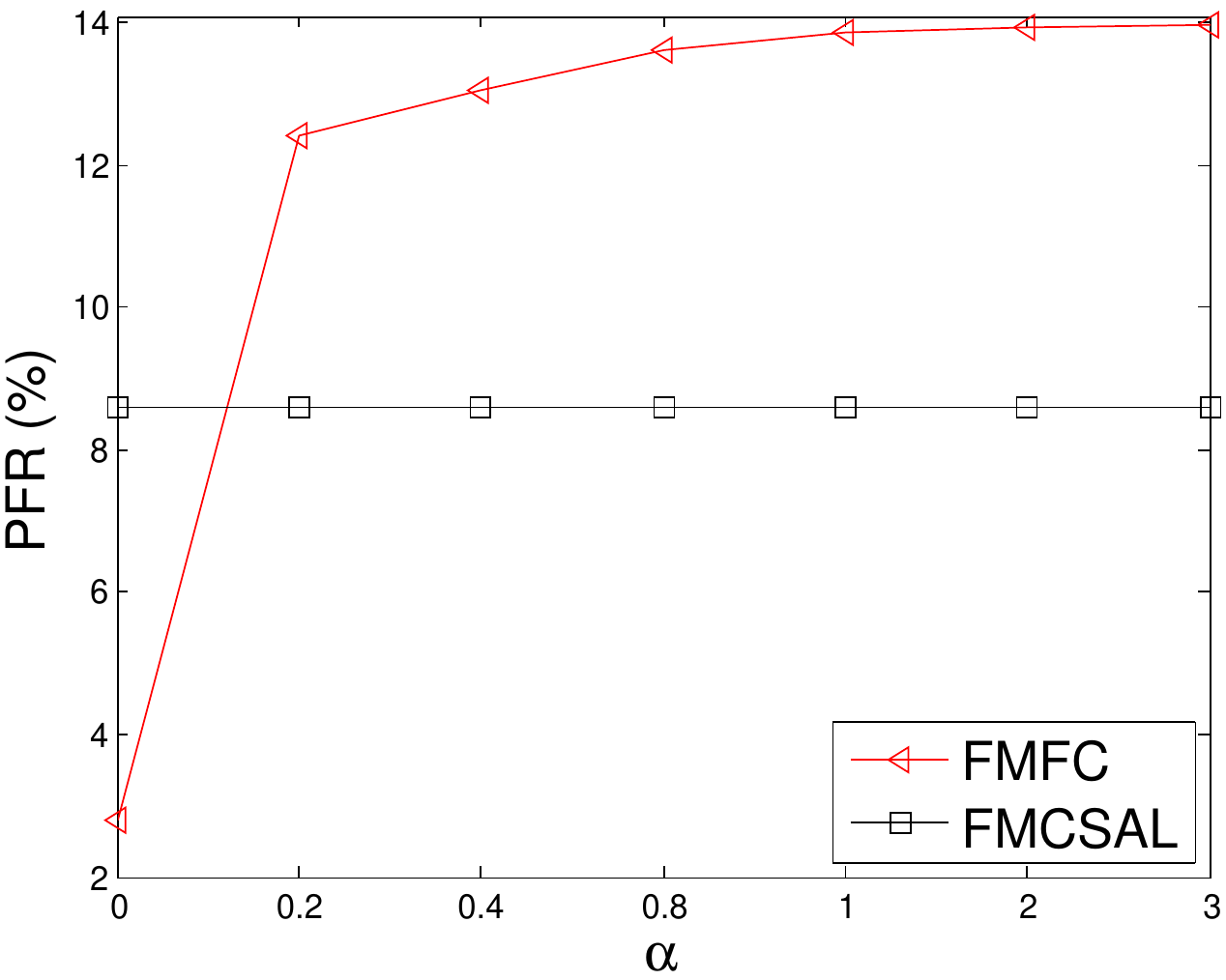}}
\end{center}
\vspace*{-10pt}
   \caption{Performance of FMFC and FMCSAL (the best compared method) measured by $RMSE$ (for the prediction phase) and $PFR$ (for the active learning phase) when we vary $\alpha$.
}
\label{fig:performance vs parameters}
\vspace*{-10pt}
\end{figure}

Results in Table \ref{table: performance comparision Movielens-IMDB} and Table \ref{table: performance comparision Amazon} have shown that our method can achieve high performance for both \emph{exploitation} (high $PFR$ in the active learning phase) and \emph{exploration} (low $RMSE$ and $MAE$ in the prediction phase). Now we analyze how $\alpha$, i.e. the weight of Criterion (1), can further balance their trade-off. We vary $\alpha$ but fix other tuned parameters to see how the performance changes. As shown in Figure \ref{fig:performance vs parameters}, $RMSE$ of FMFC first decreases then increases when $\alpha$ varies, and obtains the best result when $\alpha$ is around $1$. $PFR$ keeps increasing when $\alpha$ varies. We certainly will not set $\alpha < 1$, which will achieve poor performance in terms of both $RMSE$ and $PFR$. For $\alpha \geq 1$, if we attach more importance to the active learning phase, we need to assign a larger value to $\alpha$. Similarly, if we pay more attention to the prediction phase, then a smaller value is assigned. These experiment results are consistent with the analysis of section 4.4.

\section{Conclusion}
In this paper, we propose a novel recommendation scheme for the item cold-start problem by leveraging both active learning and items' attribute information. We firstly pre-train the rating prediction model with users' historical ratings and items' attributes. Secondly, given a new item, a small portion of users are selected to rate this item based on four useful criteria. Thirdly, the prediction model is re-trained by adding feedback ratings. Finally, unselected users' ratings are predicted by the re-trained model. We further propose a dynamic active learning budget to properly distribute active learning resources, which contributes to better recommendation performance. The idea of dynamic active learning budget can also be applied to other active learning related tasks. Our methods are able to ensure a relatively good user experience for both of selected users in the active learning phase and unselected users in the prediction phase. For future work, we will explore more other criteria to improve our user selection strategy. In addition, libFM used in this paper is a regression model, which is suitable for the task of rating prediction. We will try to expend our method with a ranking model to better address the top-$N$ recommendation task.

\section{Acknowledgement}
This work was supported in part by the National Basic Research Program of China (973 Program) under Grant 2013CB336500, National Nature Science Foundation of China (Grant Nos: 61522206, 61379071, 61373118), and National Youth Top-notch Talent Support Program.

%
%
%
%
%
%
%
%
%
%
\ifCLASSOPTIONcaptionsoff
  \newpage
\fi

\bibliographystyle{IEEEtran}
\bibliography{IEEEexample}

\begin{IEEEbiography}[{\includegraphics[width=1in,height=1.25in,clip,keepaspectratio]{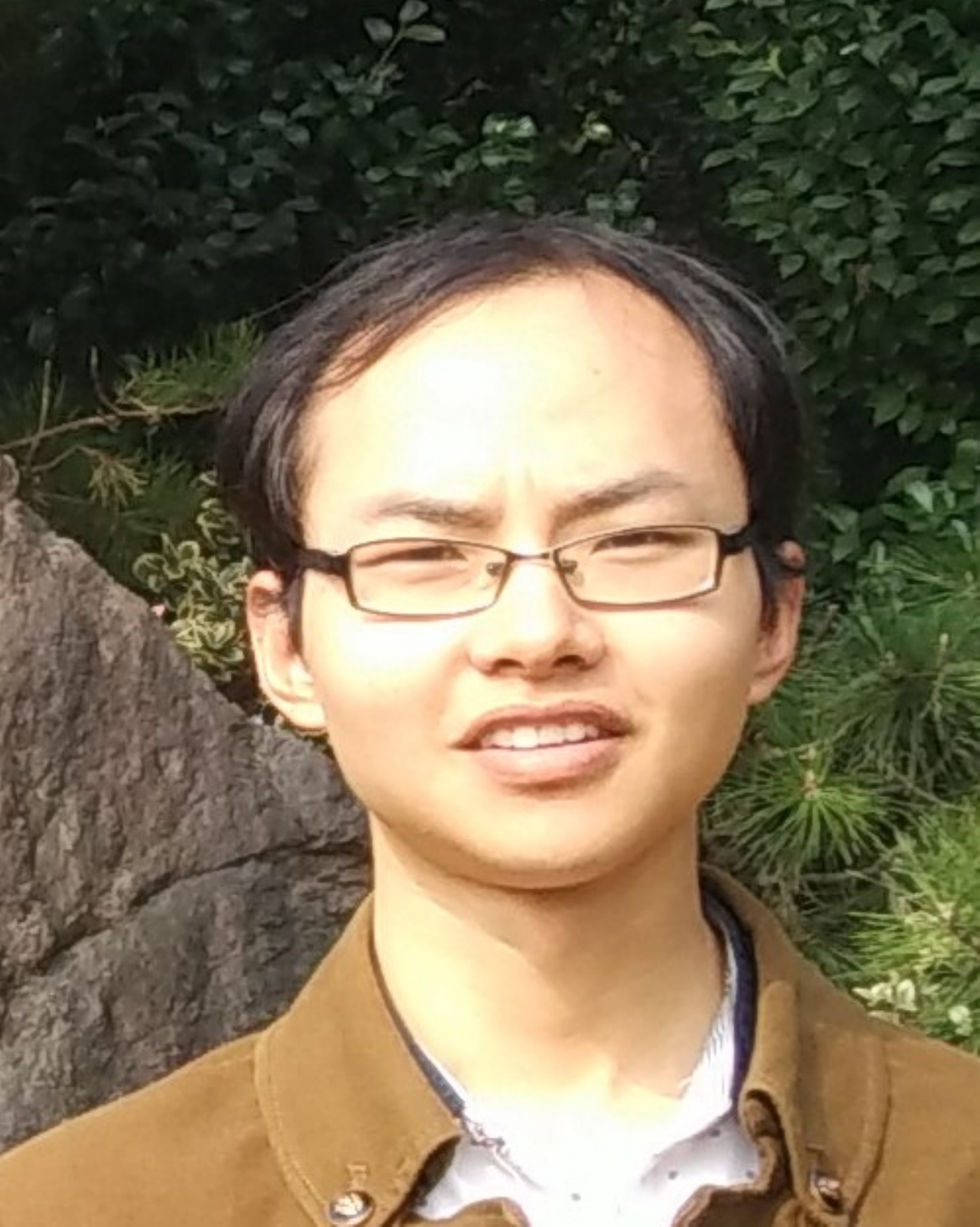}}]{Yu Zhu}
received the B.S. degree in Computer Science from Zhejiang University, China, in 2013. He is currently a Ph.D. student in computer science at Zhejiang University. His research interests include machine learning, data mining and recommender systems.
\end{IEEEbiography}
\begin{IEEEbiography}[{\includegraphics[width=1in,height=1.25in,clip,keepaspectratio]{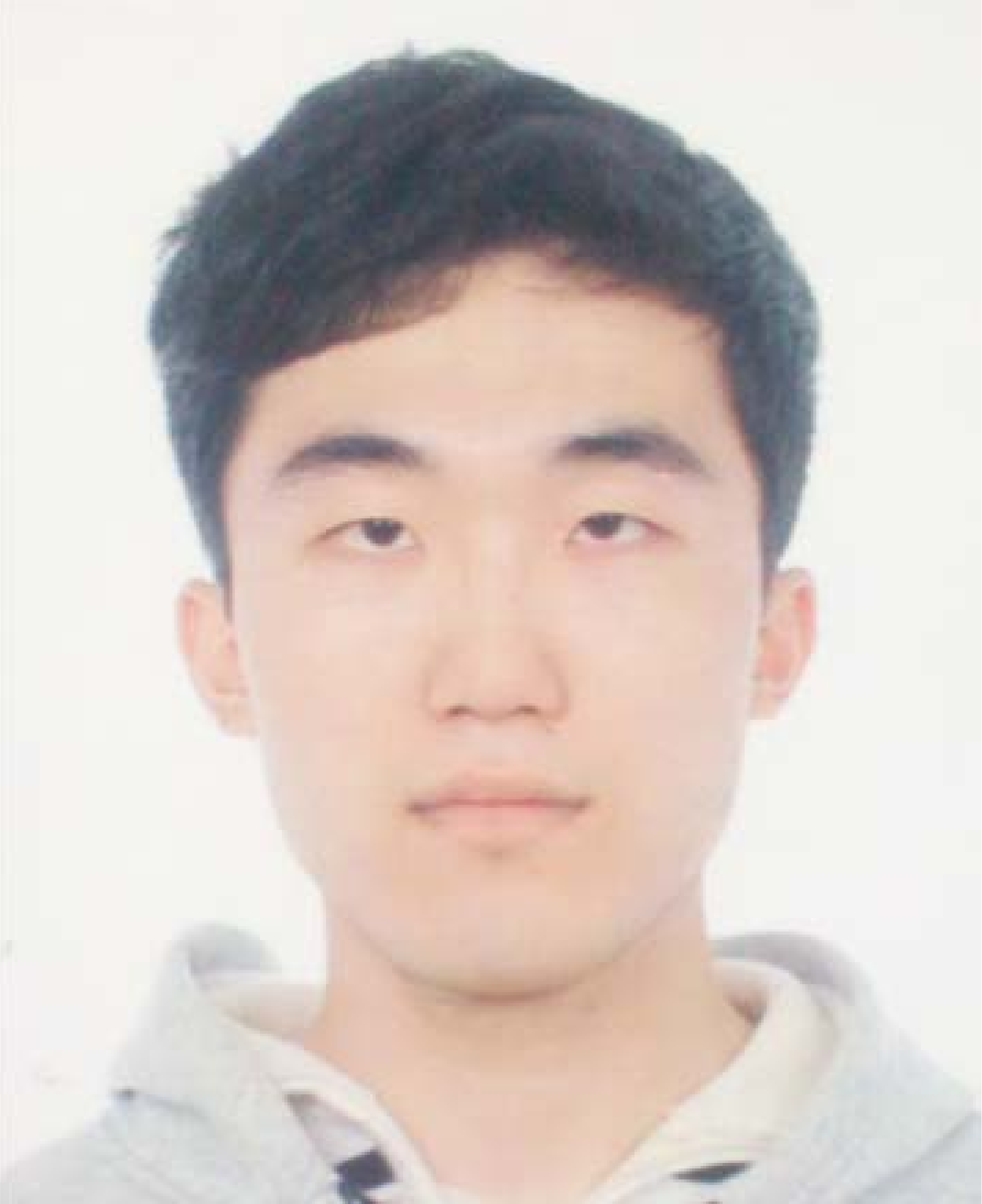}}]{Jinghao Lin}
is currently a master candidate in the State Key Lab of CAD\&CG, College of Computer Science at Zhejiang University, China. He received the BS degree from Zhejiang University, China in 2015. His research interests are data mining and recommendation systems.
\end{IEEEbiography}
\begin{IEEEbiography}[{\includegraphics[width=1in,height=1.25in,clip,keepaspectratio]{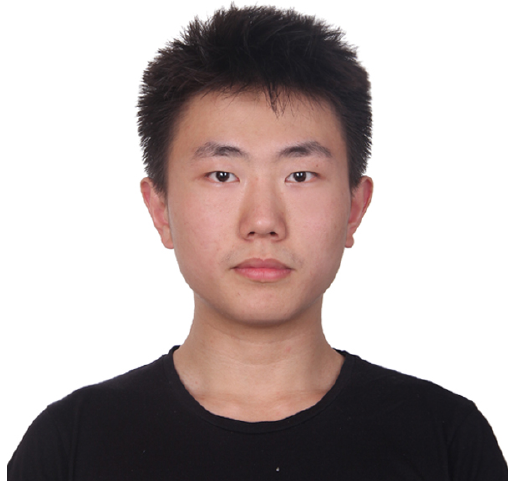}}]{Shibi He}
is a third year undergraduate in Zhejiang University. He is now a research scholar in University of Illinois at Urbana-Champaign supervised by Prof. Peng Jian. Before that, he is a member of social network group in the State Key Lab of CAD\&CG under the supervision of professor Deng Cai. His research interests include Deep Learning, Social Network, Bioinformatics and Computer Vision.
\end{IEEEbiography}
\begin{IEEEbiography}[{\includegraphics[width=1in,height=1.25in,clip,keepaspectratio]{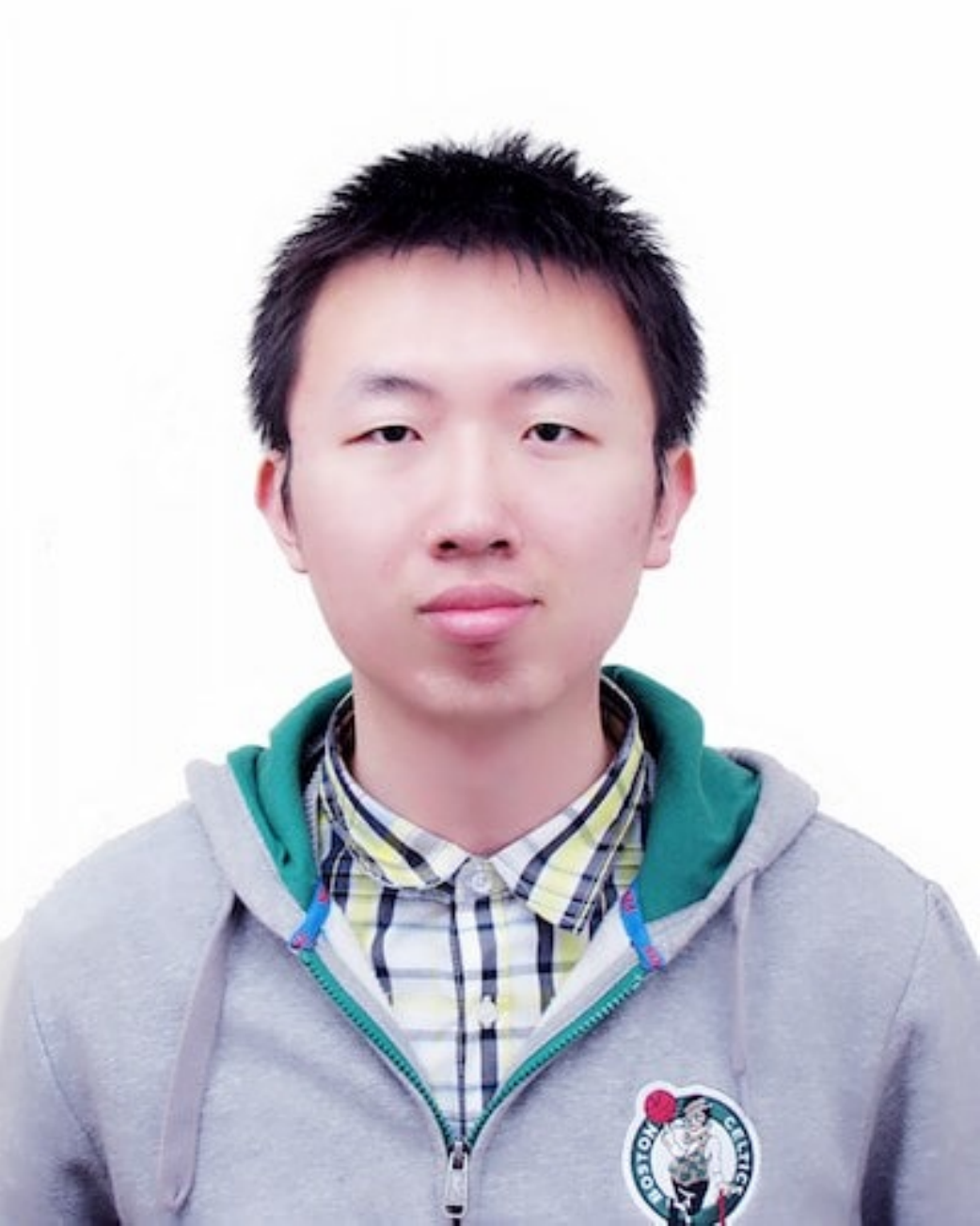}}]{Beidou Wang} received his BS degree from Zhejiang University, China, in 2011.  He is currently in a duo PhD program of Zhejiang University, China and Simon Fraser University, Canada. His research interests include social network mining and recommender systems.
\end{IEEEbiography}
\begin{IEEEbiography}[{\includegraphics[width=1in,height=1.25in,clip,keepaspectratio]{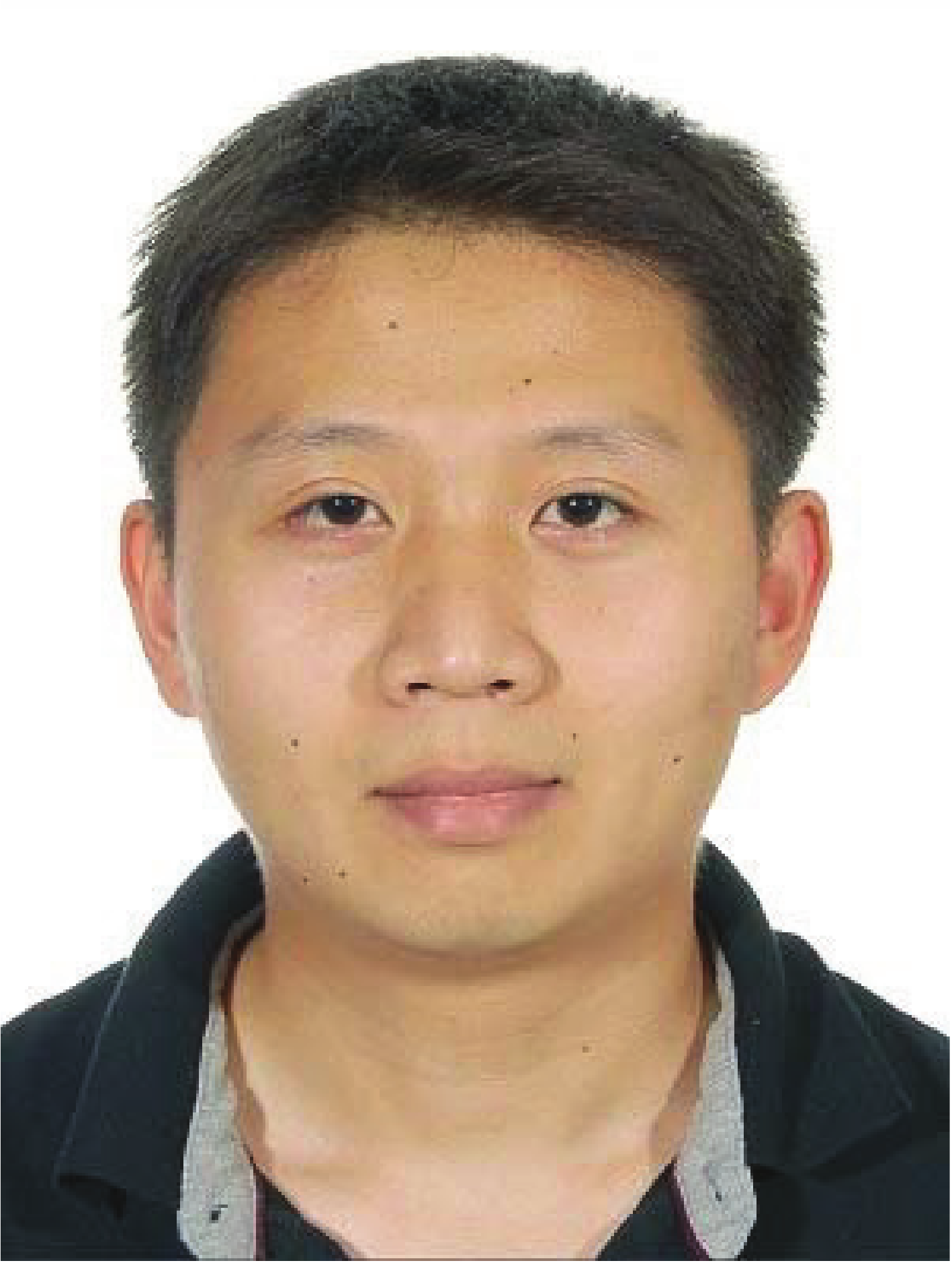}}]{Ziyu Guan}
received the BS and PhD degrees in Computer Science from Zhejiang
University, China, in 2004 and 2010, respectively. He had worked as
a research scientist in the University of California at Santa
Barbara from 2010 to 2012. He is currently a full professor in the
School of Information and Technology of Northwest
University, China. His research interests include attributed graph mining
and search, machine learning, expertise modeling and retrieval, and
recommender systems.
\end{IEEEbiography}
\begin{IEEEbiography}[{\includegraphics[width=1in,height=1.25in,clip,keepaspectratio]{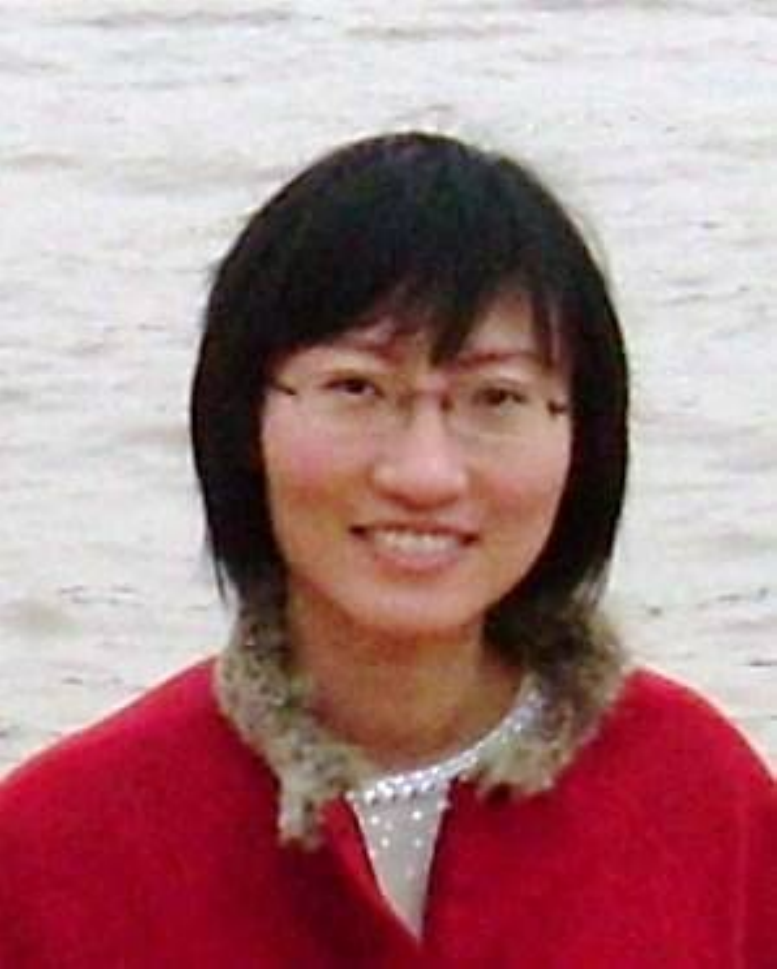}}]{Haifeng Liu} is an Associate Professor in the College of Computer Science at Zhejiang University, China. She received her Ph.D. degree in the Department of Computer Science at University of Toronto in 2009. She got her Bachelor degree in Computer Science from the Special Class for the Gifted Young at University of Science and Technology of China. Her research interests lie in the field of machine learning, pattern recognition, and web mining.
\end{IEEEbiography}
\begin{IEEEbiography}[{\includegraphics[width=1in,height=1.25in,clip,keepaspectratio]{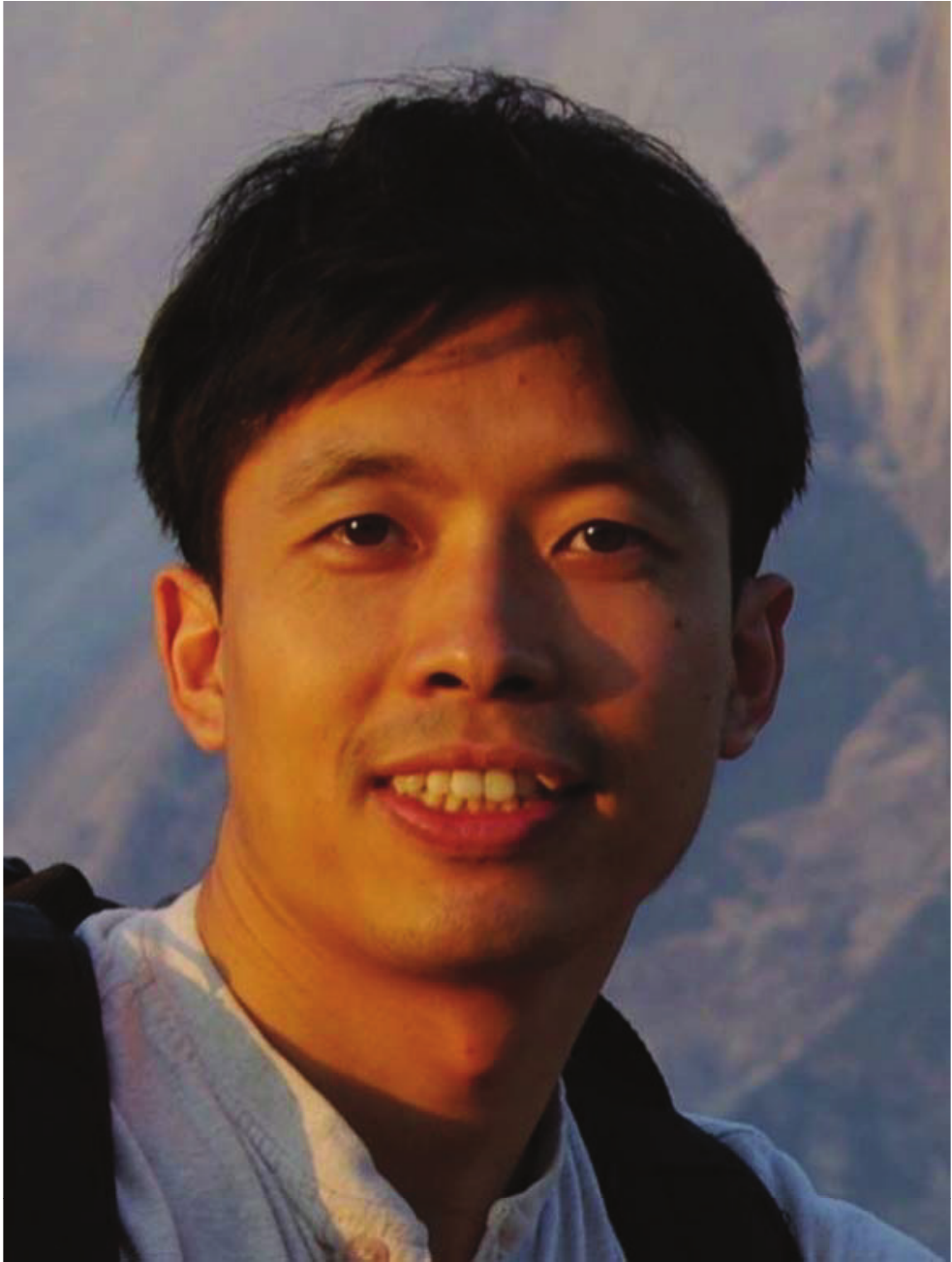}}]{Deng Cai}
is a Professor in the State Key Lab of CAD\&CG, College
of Computer Science at Zhejiang University, China. He received the
PhD degree in computer science from University of Illinois at Urbana
Champaign in 2009. Before that, he received his Bachelor's degree
and Master's degree from Tsinghua University in 2000 and 2003
respectively, both in automation. His research interests include
machine learning, data mining and information retrieval. He is a
member of the IEEE.
\end{IEEEbiography}

\end{document}